\pdfoutput=1
\documentclass[11pt]{article}

\usepackage{latex/acl}
\usepackage{xcolor}
\usepackage{graphicx}
\usepackage{subfigure}
\usepackage{stfloats}
\usepackage{color}
\usepackage{pifont}
\usepackage{balance}
 
\definecolor{midnightgreen}{rgb}{0.0, 0.29, 0.33}

\usepackage{times}
\usepackage{latexsym}
\usepackage{amsmath}
\usepackage[T1]{fontenc}
\usepackage[utf8]{inputenc}
\usepackage{amssymb}
\usepackage{microtype}

\usepackage{inconsolata}
\usepackage{multirow}
\usepackage{graphicx}
\title{{\bf \textsc{INTERVENOR}\includegraphics[width=1.1em]{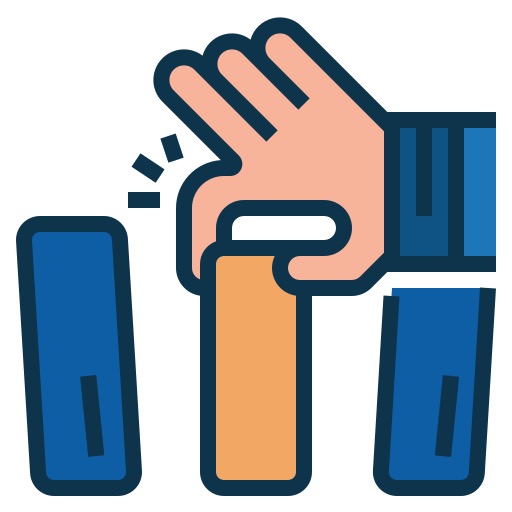}}: Prompting the Coding Ability of Large Language Models with the Interactive Chain of Repair}

\author{Hanbin Wang$^{1}$, Zhenghao Liu$^{1}$\thanks{ \ \ indicates corresponding author.}, Shuo Wang$^{2}$, Ganqu Cui$^{2}$, Ning Ding$^{2}$, \\ \textbf{Zhiyuan Liu$^{2}$ and Ge Yu$^{1}$} \\ 
$^1$Department of Computer Science and Technology, Northeastern University, China \\
$^2$Department of Computer Science and Technology, Institute for AI, Tsinghua University, China \\
Beijing National Research Center for Information Science and Technology, China \\
}
\begin{document}
\maketitle
\begin{abstract}
This paper introduces \textbf{INTERVENOR} (INTERactiVE chaiN Of Repair), a system designed to emulate the interactive code repair processes observed in humans, encompassing both code diagnosis and code repair. INTERVENOR prompts Large Language Models (LLMs) to play distinct roles during the code repair process, functioning as both a \texttt{Code Learner} and a \texttt{Code Teacher}. Specifically, the \texttt{Code Learner} is tasked with adhering to instructions to generate or repair code, while the \texttt{Code Teacher} is responsible for crafting a Chain-of-Repair (CoR) to serve as guidance for the \texttt{Code Learner}. During generating the CoR, the \texttt{Code Teacher} needs to check the generated codes from \texttt{Code Learner} and reassess how to address code bugs based on error feedback received from compilers. Experimental results demonstrate that INTERVENOR surpasses baseline models, exhibiting improvements of approximately 18\% and 4.3\% over GPT-3.5 in code generation and code translation tasks, respectively. Our further analyses show that CoR is effective to illuminate the reasons behind bugs and outline solution plans in natural language. With the feedback of code compilers, INTERVENOR can accurately identify syntax errors and assertion errors and provide precise instructions to repair codes. All data and codes are available at \url{https://github.com/NEUIR/INTERVENOR}.

% All data and codes will be released via GitHub.

\end{abstract}

\section{Introduction}
Large Language Models (LLMs), such as ChatGPT~\cite{chatgpt}, have shown remarkable performance on code related tasks~\cite{openai2023gpt4,roziere2023code,wang2023codet5+}. This has significantly enhanced the efficiency and productivity in coding and software development~\cite{qian2023communicative}. Current approaches for code-based models involve pretraining language models on code corpora~\cite{muennighoff2023octopack,luo2023wizardcoder,li2023starcoder,zheng2023codegeex} and employing Chain-of-Thought (CoT) to prompt the coding proficiency of LLMs~\cite{wei2023chainofthought,huang2023codecot,li2023structured}. However, compelling LLMs to directly generate entirely correct code proves to be exceptionally challenging, even for proficient programmers in real-world scenarios~\cite{chen2023teaching}.

\begin{figure}[t] \centering
    \includegraphics[width=0.48\textwidth]{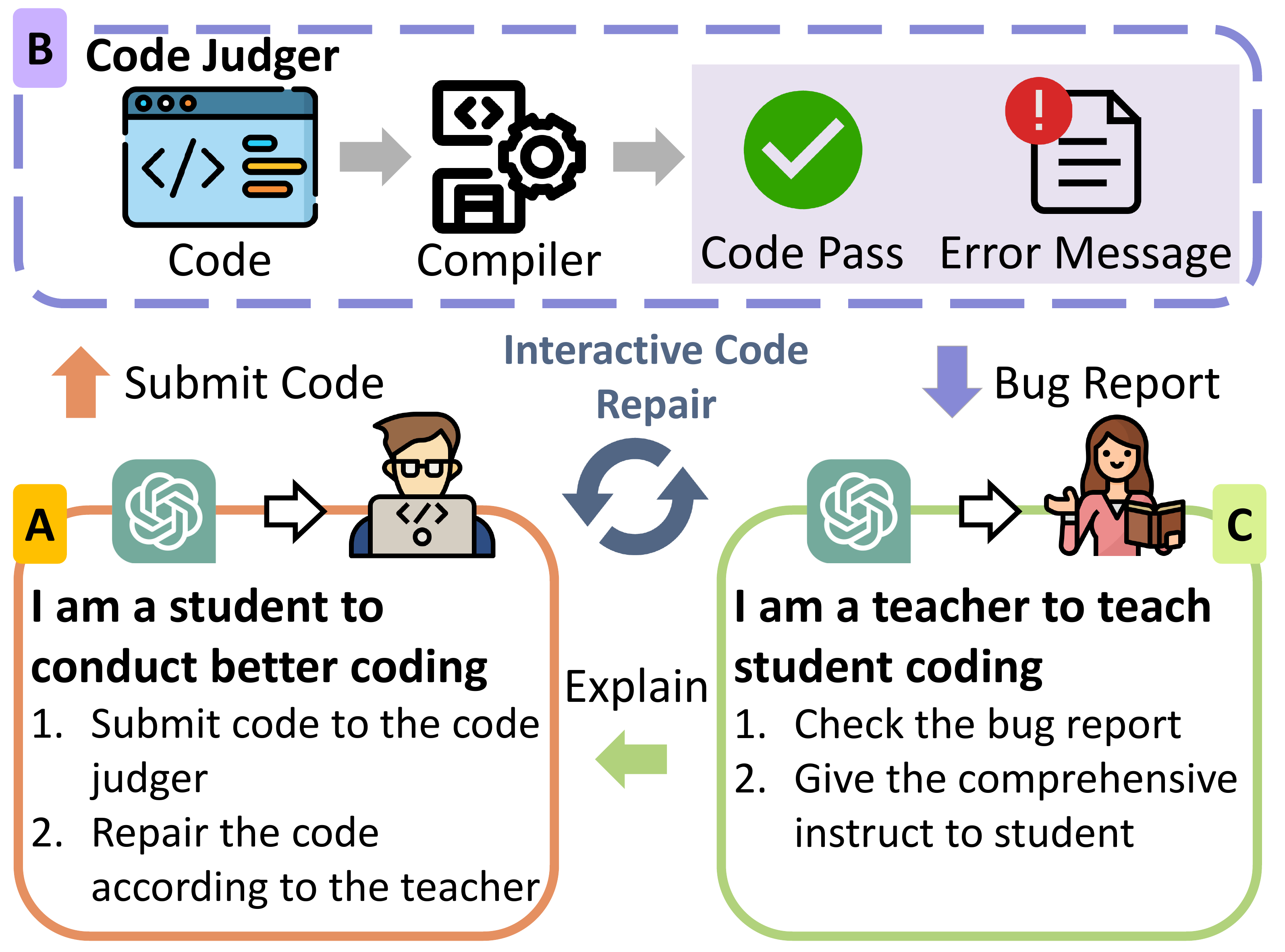}
    \caption{The Illustration of INTERVENOR. There are two agents in INTERVENOR, the teacher and student, who collaborate to repair the code. The error messages are utilized as a kind of INTERVENOR\includegraphics[width=1.1em]{figures/image/title.jpg} to alleviate the Degeneration-of-Thought (DoT) problem.} \label{fig:intro}
\end{figure}
Recently, researchers focus on improving the code generation ability of LLMs through Self-Repair techniques~\cite{Olausson2023IsSA,chen2023teaching}. These methods leverage LLMs themselves to execute and repair codes, thereby enhancing the quality of generated code. Moreover, multi-agent collaborative coding approaches~\cite{qian2023communicative,dong2023selfcollaboration} have also proven their effectiveness in handling difficult code tasks by prompting LLMs to play different roles, such as developers and testers. However, lots of bugs are difficult to find due to the cognitive inertia~\cite{mcguire1960cognitive}--overlooking the buggy codes that may not conform to their pre-existing coding thinking of LLMs. These agent-based code refinement methods~\cite{dong2023selfcollaboration,qian2023communicative} heavily rely on the self-evaluation capabilities of LLMs, potentially encountering the Degeneration-of-Thought (DoT) problem~\cite{liang2023encouraging,shinn2023reflexion}.

This paper proposes \textbf{INTER}acti\textbf{VE} chai\textbf{N} \textbf{O}f \textbf{R}epair (INTERVENOR) to alleviate the DoT problem in code repair. The approach incorporates feedback from code compilers to enhance the code repair process. Following~\citet{dong2023selfcollaboration}, we develop two agents, namely, \texttt{Code Learner} and \texttt{Code Teacher}, to collaboratively repair codes interactively. As illustrated in Figure~\ref{fig:intro}, the \texttt{Code Learner} is tasked with generating/repairing codes based on provided instructions. To craft specific code repair instructions to guide the \texttt{Code Learner}~\cite{kaddour2023challenges,wang2022self}, the \texttt{Code Teacher} generates the Chain-of-Repair (CoR) to illustrate the bug repair solutions for the \texttt{Code Learner}. Instead of Self-Debug~\cite{chen2023teaching}, \texttt{Code Teacher} incorporates the bug report from compilers to rethink the reasons of code errors and generate planning on how to repair the bugs. This interactive code repair process will continue until the code learner successfully fixes all code errors or reaches the predetermined maximum number of repair attempts.

Experimental results demonstrate the effectiveness of INTERVENOR by outperforming previous baseline models. Notably, INTERVENOR also achieves about 18\% and 4.3\% improvements over GPT-3.5~\cite{chatgpt} in code generation and code translation tasks, showing its ability to improve the quality of generated codes through iterative code repair. Besides, we also build the CodeError dataset for evaluating the code repair ability of INTERVENOR by collecting the buggy code snippets from GPT-3.5 and real-world user-submitted codes. INTERVENOR further validates its efficacy by demonstrating a twofold increase in the number of successfully repaired codes.

Our further analyses illustrate that INTERVENOR is effective in leveraging the bug messages from code compilers, recognizing the reasons for code errors, and providing correction planning in natural language. Thanks to our CoR mechanism, INTERVENOR avoids thinking by LLMs themselves and can accurately diagnose the buggy codes and correct the assertion errors and name/syntax errors even in more difficult code generation scenarios. Our CoR mechanism enables LLMs to avoid designing complex code generation/repair prompts and achieve the best performance via only three-turn code repair. It also shows the potential to leverage the feedback from environments or rule systems to evolve LLMs~\cite{Olausson2023IsSA}.

\section{Related Work}
Code generation tasks~\cite{chen2021evaluating,austin2021program,zheng2023codegeex} aim to generate correct and executable code based on the given natural language description, which has drawn lots of attention from researchers. LLMs such as ChatGPT~\cite{chatgpt} and GPT-4~\cite{openai2023gpt4} have shown strong effectiveness in generating code of high quality. To enhance the coding ability of LLMs, existing work focuses on code-specific pretraining and performs exceptionally well in code generation tasks~\cite{roziere2023code,li2023starcoder,luo2023wizardcoder,wang2023codet5+}.

Recently, some models focus on utilizing prompting techniques to enhance the coding capabilities of LLMs. CodeCoT~\cite{huang2023codecot} is inspired by Chain-of-Thought~\cite{wei2023chainofthought} and prompts the quality of generated codes using Code-CoT and Self-exam methods. LLMs are asked to craft the code and design a set of test cases to polish the codes. Structured Chain-of-Thought (SCoT)~\cite{li2023structured} further considers the program structure, such as sequences, branches, and loops, and prompts LLMs to generate intermediate reasoning steps with program structures. Nevertheless, forcing LLMs to directly generate the completely correct codes is challenging in the code generation task~\cite{chen2023teaching}.

To generate more accurate code, existing efforts primarily concentrate on Self-Refine~\cite{Olausson2023IsSA} and Self-Repair~\cite{chen2023teaching} techniques. The Self-Refine models aspire to improve the quality of generated code by decoding multiple samples and subsequently selecting the most suitable one based on specific criteria. One strategy to formulate customized criteria involves executing the generated code and selecting the optimal one based on the resulting execution outcomes~\cite{ni2023lever, zhang2022coder, shi2022natural, Li_2022}. Another approach is to rerank multiple code solutions to determine the final code~\cite{shi2022natural, zhang2022coder, chen2022codet, inala2022faultaware}. However, these methods necessitate significant computing resources for generating code candidates, rendering them inefficient~\cite{zhang2023selfedit}.

Another research avenue involves employing an iterative code repair approach to enhance the quality of generated code~\cite{zhang2023selfedit,welleck2022generating,madaan2023selfrefine,shinn2023reflexion,josifoski2024flows}. Self-Debug~\cite{chen2023teaching} utilizes explanations generated by LLMs to rectify self-generated code, while Self-Repair~\cite{Olausson2023IsSA} incorporates human-provided feedback for improvement. Self-Edit~\cite{zhang2023selfedit} employs error messages to refine generated code, but it necessitates the training of an additional fault-aware editor to generate a new program. Instead of directly integrating feedback for code repair, INTERVENOR designs an additional agent to reflect the reasons for coder errors and generate the Chain-of-Repair (CoR). This CoR is then employed to instruct the other agent in code repair through natural language.

Moreover, the work~\cite{dong2023selfcollaboration,qian2023communicative} also designs a multi-agent collaborative approach to simulate the software development process and improve the efficiency of code generation. Nevertheless, these methods are highly dependent on the self-evaluation ability of LLMs and may face the Degeneration-of-Thought (DoT) problem~\cite{liang2023encouraging,shinn2023reflexion}. 
Unlike them, INTERVENOR focuses on the bug-fixing process and proposes a simple but effective solution, which utilizes external tools, such as the Python interpreter, to execute the code~\cite{xu2023lemur,qian2023creator} and use the accurate bug report to facilitate the agent collaboration during interactive code repair.

\begin{figure*}[t] \centering
    \includegraphics[width=1\textwidth]{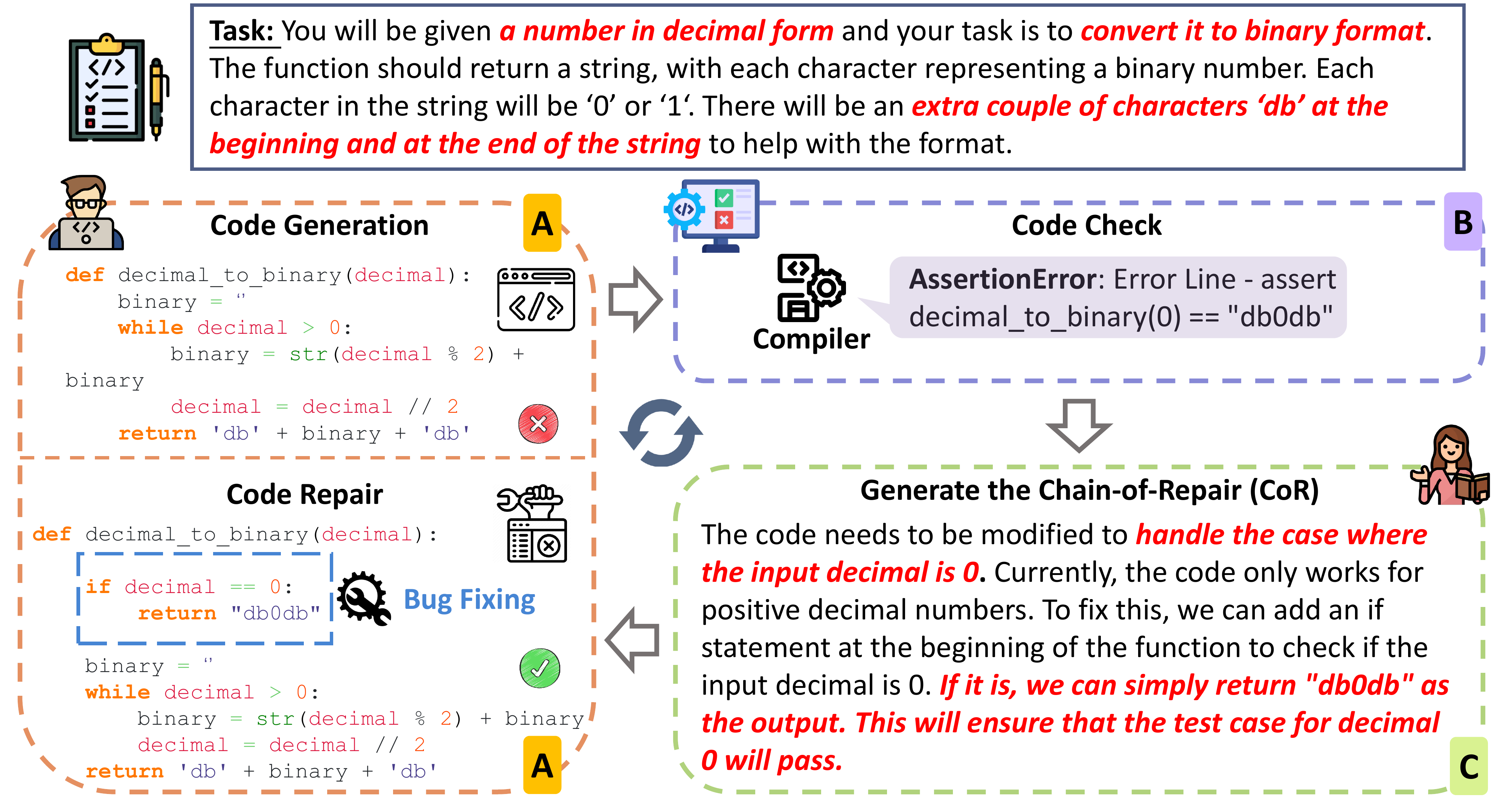}
    \caption{Illustration of Our Interactive Chain-of-Repair Model (INTERVENOR {\includegraphics[width=1.2em] {figures/image/title.jpg}}).} \label{fig:model}
\end{figure*}
\section{Methodology}

In this section, we introduce INTERVENOR, which conducts an interactive program repair process using LLM collaboration. We first describe the preliminary of code repair (Sec.~\ref{model:pre}) and then introduce our interactive Chain-of-Repair (CoR) mechanism (Sec.~\ref{model:codeagent}).

\subsection{Preliminary of Code Repair}\label{model:pre}
The code repair models mainly focus on Self-Repair~\cite{Olausson2023IsSA,chen2023teaching}. These models usually consist of three steps, including code generation, code execution, and code explanation. Self-Repair aims to use LLM itself to conduct the self-debug and self-execution processes and then iteratively repair codes. Nevertheless, programmers usually fail to recognize the code errors because of cognitive inertia~\cite{mcguire1960cognitive}, making the code execution more difficult by LLM itself.

Different from these self-repair models, INTERVENOR follows previous work~\cite{zhang2023selfedit,wang2022compilable} to incorporate the feedback from compilers to prompt the code generation ability of LLMs. Instead of directly feeding the code bug message to LLMs, we design the \textit{interactive chain of repair mechanism}, which builds two agents to rethink and repair the code errors. The compiler serves as an INTERVENOR to avoid the Degeneration-of-Thought (DoT) problem during the interactive repair process.

\subsection{Interactive Chain-of-Repair (CoR)} \label{model:codeagent}
As shown in Figure \ref{fig:model}, given the code generation tasks, INTERVENOR aims to mimic the human bug-repairing behavior by iteratively acquiring feedback from compilers, then generating bug reasons and solving plans in natural language, and finally fixing the program. Specifically, INTERVENOR employs two LLM based agents to play different roles in code repair (Sec.~\ref{sec:agent}). Then we conduct an interactive code repair process using the agents (Sec.~\ref{sec:workflow}). 

\subsubsection{Agent Building}~\label{sec:agent}
INTERVENOR involves the integration of two agents, \texttt{Code Learner} and \texttt{Code Teacher}, who work interactively to repair the generated codes. The role of the \texttt{Code Learner} follows the instructions to conduct code generation/repair, guided by the \texttt{Code Teacher}. The primary focus of the \texttt{Code Teacher} is to rethink and elucidate code errors for students. Additional details about agent construction can be found in Appendix~\ref{app:prompt}.

\textbf{Code Learner.} 
The \texttt{Code Learner} follows instructions and engages in two coding tasks, including initial code generation and code repair.

In the initial code generation phase, the \texttt{Code Learner} endeavors to generate the initial version of code according to the requirements of the given coding task. Subsequently, the agent's role is modified for code repair. We trigger the code repair proficiency of LLMs using the instruction ``You are a student assistant with excellent code repair capabilities''. Besides, \texttt{Code Learner} incorporates the Chain-of-Repair (CoR) as an instruction to guide the code repair.

\textbf{Code Teacher.} The \texttt{Code Teacher} conducts the CoR results during interactive code repair. It aims to generate the code error explanation and give bug-fixing planning for \texttt{Code Learner}.

Specifically, we use the prompt--``You are an experienced and insightful programming instructor''--to instruct LLMs to function as proficient code debuggers and serve to activate their bug-tracing and code-diagnosis abilities. \texttt{Code Teacher} integrates feedback from code compilers to produce extensive repair suggestions and guidance. This assists \texttt{Code Learner} in gaining a deeper understanding and effectively addressing errors within their code.

\subsubsection{Interactive Code Repair Workflow}~\label{sec:workflow}
INTERVENOR conducts an interactive code-repair process, facilitating the collaboration among agents and the code compiler.

In the initial step (\textbf{Step A$^0$}, where $0$ signifies the initial turn), we prompt the \texttt{Code Learner} to generate code for the given task. Subsequently, the \texttt{Code Learner} executes the generated code using the code compiler to assess its correctness (\textbf{Step B}). Following this, the \texttt{Code Teacher} generates code repair instructions (CoR) based on the bug report and the associated buggy code (\textbf{Step C}). These instructions elucidate the reason of the bug, such as ``modified to handle the case where the input decimal is 0'', and include code correction planning, for example, ``we can simply return db0db''. Such instructions are informative and enhance the guidance for the \texttt{Code Learner}. Ultimately, the \texttt{Code Learner} follows the \textit{chain-of-repair (CoR)} to rectify the code and subsequently resubmits the corrected version to the compiler for execution in the subsequent turn ((\textbf{Step A$^i$}), where $i\geq1$ denotes the code repair process). The A$^i$, B, and C steps are iterated sequentially until either the code meets the compiler's estimation or the maximum turn limit is reached.

\section{Experimental Methodology}
In this section, we describe the datasets, evaluation metrics, baselines, and implementation details.

\begin{table}[]
\small
\centering
\begin{tabular}{llrr}
\hline
\textbf{Benchmark}           & \textbf{Language} & \textbf{Problems} & \textbf{\#Tests} \\ \hline
HumanEval                    & Python            & 164               & 7.8                \\ \hline
MBPP                         & Python            & 500               & 3.1                \\ \hline
\multirow{3}{*}{HumanEval-X} & C++               & 164               & 7.8                \\
                             & Java              & 164               & 7.8                \\
                             & JavaScript        & 164               & 7.8                \\ \hline
CodeError                    & Python            & 4,463               & 9.0                \\ \hline
\end{tabular}
\caption{Data Statistics. \#Tests represents the average number of test cases.}
\label{tab:data_statistic}
\end{table}

\textbf{Dataset.} We evaluate the code generation and translation effectiveness on three datasets, including \texttt{HumanEval}, \texttt{MBPP} and \texttt{HumanEval-X}. Besides, we build a new benchmark \texttt{CodeError} to further test the code repair ability of LLMs. All data statistics are shown in Table~\ref{tab:data_statistic}.

\texttt{HumanEval (HEval)}~\cite{chen2021evaluating} serves as a benchmark for evaluating the functional correctness of synthesized programs generated from docstrings. It comprises 164 hand-written Python programming problems, which consist of function signatures, docstrings, bodies, and multiple unit tests. \texttt{MBPP}~\cite{austin2021program} is a benchmark that includes 974 introductory-level Python programming problems. Each problem comprises a problem statement, a code solution, and three automated test cases, and the task IDs that range from 11 to 510 are used for evaluation.
\texttt{HumanEval-X}~\cite{zheng2023codegeex} is used to assess a model's multi-programming language generation and translation capability. It consists of 820 human-crafted data instances, covering C++, Java, JavaScript, and Go. 
\texttt{DS-1000}~\cite{Lai2022DS1000} is a benchmark designed to evaluate the capabilities of LLMs in data science code generation. It includes 1,000 problems that span seven Python libraries, such as NumPy and Pandas. We use the completion style prompt for each question.

Then we build the \texttt{CodeError} benchmark to further evaluate the code repair effectiveness of INTERVENOR. The CodeError benchmark contains a total of 4,463 examples, evenly distributed across more than six different error types. It includes basic programming problems, data analysis problems, and programming competition problems. For each example, there are 9 test cases on average to evaluate the code's correctness. More details of \texttt{CodeError} are shown in Appendix~\ref{app:codeerror}.

\textbf{Evaluation Metrics. } We use Pass@$k$~\cite{chen2021evaluating} to evaluate the effectiveness of different models on both code generation task and code translation task, which is the same as the previous work~\cite{chen2021evaluating,zheng2023codegeex,li2023structured,chen2022codet,nijkamp2022codegen}.

\textbf{Baselines.} We first compare INTERVENOR with several code-oriented large language pretrained models, such as Incoder~\cite{fried2022incoder}, CodeGen~\cite{nijkamp2022codegen}, CodeGeeX~\cite{zheng2023codegeex}, CodeT5~\cite{wang2023codet5+}, StarCoder~\cite{li2023starcoder}, WizardCoder~\cite{luo2023wizardcoder}, and Llama based models~\cite{touvron2023llama,roziere2023code}. These models are pretrained on large-scale code corpora, demonstrating strong code generation capabilities. Additionally, we also compare INTERVENOR with some closed-source and high-performance large language models, \textit{e.g.} Claude~\cite{Claude}, GPT-3.5~\cite{chatgpt}, and GPT-4~\cite{openai2023gpt4}, which show strong emergent abilities, especially for the code generation tasks. In our experiments, GPT-3.5 is our main baseline model. Besides, we also compare Self-Debug~\cite{chen2023teaching} and the multi-agent collaborative method, Self-Collaboration~\cite{dong2023selfcollaboration} to show the code repair ability of INTERVENOR.

\textbf{Implementation Details.} In our experiments, we use GPT-3.5 (gpt-3.5-turbo-0613) as the foundation model to build different agents in INTERVENOR. We set the temperature to 0.2 and the maximum generation length to 512 tokens. The maximum number of interactive code repairs is set to 5. Additionally, we also use CodeLlama-7B/13B-Instruct to implement the agents, \texttt{Code Learner} and \texttt{Code Teacher}, of our INTERVENOR model to explore the impact of using different LLMs. On all datasets, we use the 0-shot setting in our experiments.

\begin{table*}[t]
\centering
\small
\resizebox{\linewidth}{!}{
\begin{tabular}{l|cccccc|cccc}
\hline
\multirow{3}{*}{\textbf{Model}} & \multicolumn{6}{c|}{\textbf{Code Generation}}                                                                                                                        & \multicolumn{4}{c}{\textbf{Code Translation}}                 \\ \cline{2-11} 
                                & \multicolumn{1}{c|}{\textbf{HEval}} & \multicolumn{1}{c|}{\textbf{MBPP}} & \multicolumn{3}{c|}{\textbf{HumanEval-X}}                          & \textbf{DS-1000} & \multicolumn{4}{c}{\textbf{Target Language}}                  \\
                                & \multicolumn{1}{c|}{Python}             & \multicolumn{1}{c|}{Python}        & C++           & Java          & \multicolumn{1}{c|}{JS}            & Python           & Python        & C++           & Java          & JS            \\ \hline
InCoder~\cite{fried2022incoder}                         & \multicolumn{1}{c|}{15.2}               & \multicolumn{1}{c|}{19.4}          & 9.5           & 9.1           & \multicolumn{1}{c|}{13.0}          & 7.4              & -             & -             & -             & -             \\
CodeGen~\cite{nijkamp2022codegen}                         & \multicolumn{1}{c|}{18.3}               & \multicolumn{1}{c|}{20.9}          & 18.1          & 14.9          & \multicolumn{1}{c|}{18.4}          & 8.4              & 40.7          & 37.6          & 35.4          & 51.8          \\
CodeGeeX~\cite{zheng2023codegeex}                        & \multicolumn{1}{c|}{22.9}               & \multicolumn{1}{c|}{24.4}          & 17.1          & 20.0          & \multicolumn{1}{c|}{17.6}          & -                & 68.5          & 43.6          & 56.8          & 45.2          \\
CodeT5+~\cite{wang2023codet5+}                         & \multicolumn{1}{c|}{30.9}               & \multicolumn{1}{c|}{-}             & -             & -             & \multicolumn{1}{c|}{-}             & -                & -             & -             & -             & -             \\
InstructCodeT5+~\cite{wang2023codet5+}                 & \multicolumn{1}{c|}{35.0}               & \multicolumn{1}{c|}{-}             & -             & -             & \multicolumn{1}{c|}{-}             & -                & -             & -             & -             & -             \\
PaLM-Coder~\cite{chowdhery2022palm}                      & \multicolumn{1}{c|}{35.9}               & \multicolumn{1}{c|}{47.0}          & -             & -             & \multicolumn{1}{c|}{-}             & -                & -             & -             & -             & -             \\
StarCoder~\cite{li2023starcoder}                       & \multicolumn{1}{c|}{40.8}               & \multicolumn{1}{c|}{49.5}          & -             & -             & \multicolumn{1}{c|}{-}             & 26.0             & -             & -             & -             & -             \\
WizardCoder~\cite{luo2023wizardcoder}                     & \multicolumn{1}{c|}{57.3}               & \multicolumn{1}{c|}{51.8}          & -             & -             & \multicolumn{1}{c|}{-}             & 28.4             & -             & -             & -             & -             \\
LLama2~\cite{touvron2023llama}                          & \multicolumn{1}{c|}{30.5}               & \multicolumn{1}{c|}{45.4}          & -             & -             & \multicolumn{1}{c|}{-}             & -                & -             & -             & -             & -             \\
CodeLLama~\cite{roziere2023code}                       & \multicolumn{1}{c|}{62.2}               & \multicolumn{1}{c|}{61.2}          & -             & -             & \multicolumn{1}{c|}{-}             & 28.0             & -             & -             & -             & -             \\
PanGu-Coder2~\cite{shen2023pangucoder2}                    & \multicolumn{1}{c|}{61.6}               & \multicolumn{1}{c|}{-}             & -             & -             & \multicolumn{1}{c|}{-}             & -                & -             & -             & -             & -             \\
Claude~\cite{Claude}                          & \multicolumn{1}{c|}{47.6}               & \multicolumn{1}{c|}{-}             & -             & -             & \multicolumn{1}{c|}{-}             & -                & -             & -             & -             & -             \\
GPT-4~\cite{openai2023gpt4}                           & \multicolumn{1}{c|}{67.0}               & \multicolumn{1}{c|}{-}             & -             & -             & \multicolumn{1}{c|}{-}             & -                & -             & -             & -             & -             \\
Self-Debug (Simple)~\cite{chen2023teaching}                      & \multicolumn{1}{c|}{73.8}               & \multicolumn{1}{c|}{-}             & -             & -             & \multicolumn{1}{c|}{-}             & -                & -             & -             & -             & -             \\
Self-Debug (UT+Trace)~\cite{chen2023teaching}                      & \multicolumn{1}{c|}{71.9}               & \multicolumn{1}{c|}{-}             & -             & -             & \multicolumn{1}{c|}{-}             & -                & -             & -             & -             & -             \\
Self-Collaboration~\cite{dong2023selfcollaboration}              & \multicolumn{1}{c|}{74.4}               & \multicolumn{1}{c|}{68.2}          & -             & -             & \multicolumn{1}{c|}{-}             & -                & -             & -             & -             & -             \\ \hline
GPT-3.5~\cite{chatgpt}                         & \multicolumn{1}{c|}{60.3}               & \multicolumn{1}{c|}{39.8}          & 52.4          & 50.6          & \multicolumn{1}{c|}{54.3}          & 29.7             & 84.3          & 71.5          & 81.7          & 84.6          \\
INTERVENOR                      & \multicolumn{1}{c|}{\textbf{75.6}}      & \multicolumn{1}{c|}{\textbf{69.8}} & \textbf{67.1} & \textbf{68.3} & \multicolumn{1}{c|}{\textbf{67.1}} & \textbf{39.7}    & \textbf{89.8} & \textbf{75.6} & \textbf{85.4} & \textbf{88.3} \\ \hline
\end{tabular}
}
\caption{Overall Performance of Different Models. We evaluate model effectiveness on code generation and code translation (HumanEval-X dataset) tasks using the Pass@$1$ evaluation metric. The baseline results are borrowed from corresponding papers. Simple and UT+Trace are two variants of Self-Debug. More evaluation results on the code translation task are shown in Appendix~\ref{app:code_trans}.}
\label{tab:overall_code_gen}
\end{table*}

% The code translation experiments are conducted on HumanEval-X. 

\section{Evaluation Results}
In this section, we evaluate the overall performance of INTERVENOR. Then we conduct ablation studies and also show the effectiveness of Interactive CoR in different testing scenarios. Finally, case studies are presented.

\subsection{Overall Performance}
The overall performance of INTERVENOR in code generation and translation tasks is shown in Table~\ref{tab:overall_code_gen}. 

Overall, INTERVENOR outperforms all baselines in all tasks by achieving more than 1\% improvements, showing its effectiveness. Compared to our main baseline model GPT-3.5, INTERVENOR achieved about 18\% and  4.3\% improvements in code generation and code translation tasks, respectively. It illustrates that INTERVENOR has the ability to prompt the coding ability of LLMs by mimicking the human code repair behavior--\textit{iteratively judging, rethinking, and repairing}. Notably, INTERVENOR also surpasses Self-Debug and Self-Collaboration models, demonstrating its ability to successfully intervene in the code generation/translation process and guide LLMs to better repair the codes using our chain of repair mechanism. All these experimental results highlight the generalization ability of INTERVENOR in improving LLMs' coding ability in different languages.

% \usepackage{multirow}

% \usepackage{multirow}
\begin{table*}[t]
\centering
\small
\begin{tabular}{l|l|c|c|ccc|c|c}
\hline
\multirow{2}{*}{\textbf{Code Repair}} & \multirow{2}{*}{\textbf{Prompt Methods}} & \textbf{HEval} & \textbf{MBPP} & \multicolumn{3}{c|}{\textbf{HumanEval-X}} & \multicolumn{1}{l|}{\textbf{CodeError}} & \multirow{2}{*}{\textbf{Avg.}} \\
                          &       & Python             & Python        & C++        & Java       & JS      & Python                                  &                                   \\ \hline
\multirow{4}{*}{No Repair} & Zero-Shot            & 60.3               & 39.8          & 52.4       & 50.6       & 54.3            & -                                    & -                            \\
& Zero-Shot CoT          & 51.8               & 35.2          & 48.2       & 45.7       & 45.4            & -                                     & -                                 \\
&Few-Shot               & 62.2               & 45.4          & 53.1       & 62.2       & 43.3            & -                                     & -                             \\
& Few-Shot CoT                    & 60.4               & 45.4          & 48.2       & 63.4       & 57.9            & -                                     & -                              \\ \hline
\multirow{6}{*}{Single Turn} & Zero-Shot &62.2&41.6&54.3&65.2&60.4&4.9&48.1\\
 & Few-Shot &65.2&40.6&57.9&65.2&62.8&9.8&50.3\\
& CoT &66.5&48.8&56.7&62.2&60.4&10.3&50.8\\

& Self-Refine       & 65.2                  & 48.8             & 57.3          & 64.0          & 60.4              & 5.2                                     & 50.2                                 \\
& Error Msgs        & 67.1               & 51.8          & 57.3       & 59.8       & 62.8            & 9.8                                     & 51.4                              \\
& INTERVENOR (CoR)     & 69.5                  & 51.0             & 59.8          & 65.2          & 60.4              & 15.9                                     & 53.6                                 \\\hline
Multi-Turns & INTERVENOR  (CoR)                     & \textbf{75.6}               & \textbf{69.8}          & \textbf{67.1}       & \textbf{68.3}       & \textbf{67.1}            & \textbf{21.7}                                      & \textbf{61.6}                              \\ 
\hline
\end{tabular}
\caption{Evaluations on Different Prompting Methods. We compare INTERVENOR with different prompting techniques to evaluate its effectiveness. All models are built based on GPT-3.5 and evaluated using Pass@$1$.}
\label{tab:prompt_tec}
\end{table*}

% \usepackage{multirow}
\begin{table*}[t]
\small
\centering
\begin{tabular}{l|l|c|c|ccc|c}
\hline
\multirow{2}{*}{\textbf{Code Learner}} & \multirow{2}{*}{\textbf{Code Teacher}} & \textbf{HEval} & \textbf{MBPP} & \multicolumn{3}{c|}{\textbf{HumanExal-X}} & \textbf{CodeError} \\
                                       &                                        & Python             & Python        & C++          & Java         & JS          & Python             \\ \hline
\multirow{4}{*}{CodeLlama-7B}          & N/A                                    & 32.3               & 34.4          & 30.3         & 29.8         & 33.5        & -                  \\
                                       & CodeLlama-7B                           & 32.9               & 35.1          & 31.0         & 31.0         & 34.1        & 4.6                \\
                                       & CodeLlama-13B                          & 33.5               & 35.3          & 32.2         & 31.5         & 35.5        & 4.9                \\
                                       & GPT-3.5                                & \textbf{36.6}               & \textbf{38.4}          & \textbf{33.5}         & \textbf{32.3}         & \textbf{40.4}        & \textbf{8.6}               \\ \hline
\multirow{4}{*}{CodeLlama-13B}         & N/A                                    & 39.6               & 36.4          & 36.6         & 33.5         & 39.6        & -                  \\
                                       & CodeLlama-7B                           & 40.0               & 37.1          & 37.1         & 34.5         & 40.4        & 8.8                \\
                                       & CodeLlama-13B                          & 42.1               & 37.7          & 38.6         & 36.6         & 42.8        & 9.4                \\
                                       & GPT-3.5                                & \textbf{43.6}               & \textbf{40.4}          & \textbf{42.2}         & \textbf{37.5}         & \textbf{46.1}        & \textbf{12.6}               \\ \hline
\multirow{4}{*}{GPT-3.5}               & N/A                                    & 60.3               & 39.8          & 52.4         & 50.6         & 54.3        & -                  \\
                                       & CodeLlama-7B                           & 65.2               & 42.8          & 55.4         & 55.1         & 56.1        & 11.8               \\
                                       & CodeLlama-13B                          & 66.5               & 46.1          & 56.1         & 58.9         & 58.9        & 13.6               \\
                                       & GPT-3.5                                & \textbf{69.5}               & \textbf{51.0}          & \textbf{59.8}         & \textbf{65.2}         & \textbf{60.4}        & \textbf{15.9}               \\ \hline
\end{tabular}
\caption{Model Performance of Different \texttt{Code Learner} and \texttt{Code Teacher} Model Pairings. We only conduct one turn code repair in experiments. ``N/A'' represents the initial output of \texttt{Code Learner} without any intervention from the \texttt{Code Teacher} for code repair, it reflects the initial generation results of \texttt{Code Learner}.}
\label{tab:diff_basemodel}
\end{table*}

\subsection{Ablation Studies}
The ablation studies are conducted to show the effectiveness of the interactive CoR mechanism.

\textbf{Evaluation on Different Prompting Methods.} Firstly, we evaluate the code generation ability of LLMs using different code generation/repair prompting methods. As shown in Table~\ref{tab:prompt_tec}, we compare CoT~\cite{kojima2023large}, Few-Shot~\cite{chen2021evaluating}, and Few-Shot CoT~\cite{wei2023chainofthought} models in experiments, which prompt LLMs to better generate codes. These models try to generate natural language as the chain of coding thought (CoT) or provide some instances to demonstrate the coding task (Few-Shot). Then we compare different methods to generate the code repair instruction, including Self-Refine~\cite{madaan2023selfrefine}, Error Msgs, and Chain-of-Repair (CoR). Self-Refine asks the LLMs to rethink the errors by themselves, while Error Msgs and CoR incorporate the code error messages from compilers. Msgs directly uses error messages to guide the code repair process.

The experimental results show that code repair is more effective than directly prompting LLMs to generate codes. Even though we conduct different Few-shot and CoT methods to directly prompt LLMs, we only achieve 3.5\% improvements, which shows that it is difficult for LLMs to generate correct codes without repairing. On the contrary, the code repair methods improve the quality of generated codes by achieving more than 9.7\% improvements with only single-turn repair. Both Error Msgs and CoR thrive on the feedback from code compilers and achieve more than 1.2\% improvements than Self-Refine, demonstrating that compilers can provide valuable signals to help LLMs better recognize the code bugs. Notably, CoR achieves the best performance among all code repair models, illustrating its effectiveness in guiding LLMs for code repair. CoR uses the error messages from code compilers to prompt LLMs, aiming to rethink the reasons for making errors and generate the instructions for repairing.

\textbf{Building Code Teacher/Learner Using Different LLMs.} Then, as shown in Table~\ref{tab:diff_basemodel}, we investigate the impact of using different LLMs to build \texttt{Code Learner} and \texttt{Code Teacher}.
When we use the identical LLM to build the \texttt{Code Learner}, the performance of INTERVENOR is improved with the enhanced capabilities of the \texttt{Code Teacher}, underscoring the efficacy of our CoR mechanism. This indicates that stronger models can conduct a more in-depth analysis of erroneous code and provide more accurate suggestions for code repair. Similarly, by keeping the same LLM to build the \texttt{Code Teacher}, a stronger \texttt{Code Learner} also conducts better code repair results. This indicates that a more powerful \texttt{Code Learner} has a stronger instruction-following ability to better comprehend the CoR provided by the \texttt{Code Teacher}.

% \usepackage{multirow}
\begin{table}[t]
\centering
\small
\resizebox{1\linewidth}{!}{
\begin{tabular}{c|l|c|c|ccc|c}
\hline
\multirow{2}{*}{\textbf{\#Turn} } & \multirow{2}{*}{\textbf{Model}} & \textbf{HEval} & \textbf{MBPP}   & \multicolumn{3}{c|}{\textbf{HumanEval-X}}  & \multirow{2}{*}{\textbf{Avg.}} \\
                                    &                                 & \textbf{Python}    & \textbf{Python} & \textbf{C++} & \textbf{Java} & \textbf{JS} &                                \\ \hline
0                              & GPT-3.5                         & 60.3               & 39.8            & 52.4         & 50.6          & 54.3        & 51.5                           \\ \hline
\multirow{3}{*}{1}                     & Error Msgs                    & 67.1               & \textbf{51.8}            & 57.3         & 59.8          & \textbf{62.8}        & 59.8                           \\
& Self-Refine                     & 65.2               & 48.8            & 57.3         & 64.0          & 60.4        & 59.1                           \\
                                       & INTERVENOR                      & \textbf{69.5}               & 51.0            & \textbf{59.8}         & \textbf{65.2}          & 60.4        & \textbf{61.2}                           \\ \hline
\multirow{3}{*}{2}                     & Error Msgs                    & 68.3               & 53.0            & 60.3         & 62.8          & \textbf{63.4}        & 61.6                           \\
& Self-Refine                     & 69.5               & 49.6            & 60.3         & 64.6          & 62.8        & 61.4                           \\
                                       & INTERVENOR                      & \textbf{73.2}               & \textbf{54.4}            & \textbf{61.6}         & \textbf{67.1}          & 62.2        & \textbf{63.7}                           \\ \hline
\multirow{3}{*}{3}                     & Error Msgs                    & 69.5               & 54.9            & 62.2         & 63.7          & \textbf{64.0}        & 62.9                           \\
& Self-Refine                     & 71.9               & 51.2            & 63.4         & 65.2          & 63.4        & 63.0                           \\
                                       & INTERVENOR                      & \textbf{75.6}               & \textbf{60.2}           & \textbf{65.2}         & \textbf{68.3}          & \textbf{64.0}        & \textbf{66.7}                           \\ \hline
\end{tabular}
}
\caption{Multi-Turn Code Repair Performance of INTERVENOR, Error Msgs, and Self-Refine.}
\label{tab:dot}
\end{table}

\textbf{Effectiveness on Alleviating the DoT problem.} Finally, we conduct experiments to demonstrate how our model alleviates the Degenerate-of-Thought (DOT) problem. As shown in Table~\ref{tab:dot}, we compare the multi-turn code repair performance of INTERVENOR, Error Msgs, and Self-Refine. 
Both INTERVENOR and Error Msgs incorporate feedback from the compiler, while Self-Refine conducts self-reflection to generate the reasons for buggy code. Thus, Self-Refine usually faces the DoT problem, which usually neglects the bugs that appear in the code segments. The experiments show that with the increase in the number of repair turns, the performance of INTERVENOR surpasses that of Self-Refine and Error Msgs across almost all datasets. After three turns of code repair, INTERVENOR achieves an average improvement of 3.7\% over Self-Refine and Error Msgs, highlighting its effectiveness in alleviating the DoT problem.

\begin{figure}[t] \centering
    \includegraphics[width=0.45\textwidth]{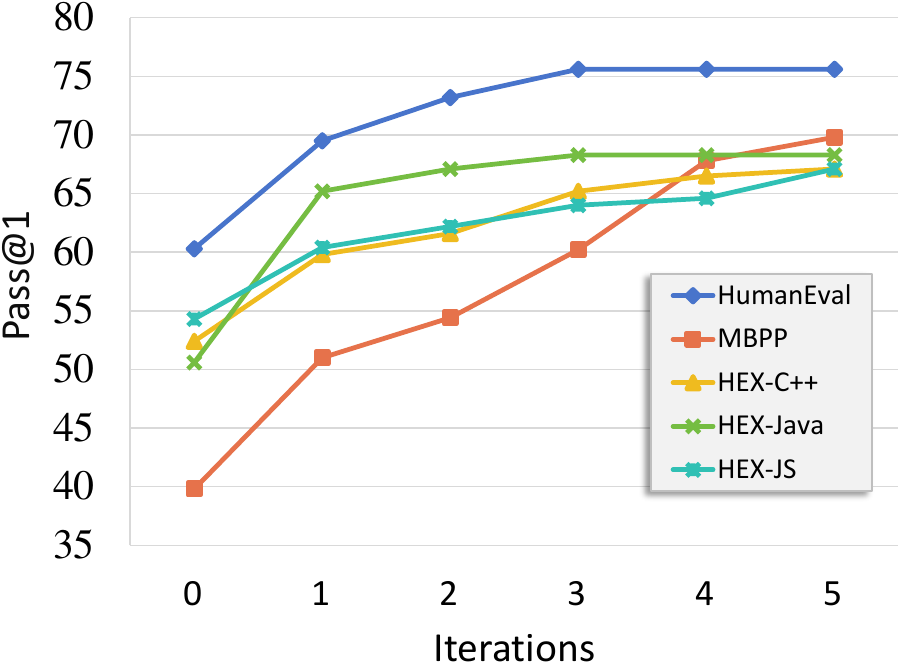}
    \caption{The Impact of Different Code Repair Turns. HumanEval, MBPP, and HumanEval-X (HEX) are used to evaluate our INTERVENOR model.} \label{fig:iterations}
\end{figure}

\subsection{Effectiveness of INTERVENOR in Different Testing Scenarios}
In this subsection, we delve deeper into exploring the effectiveness of INTERVENOR in two testing scenarios: 1) validating the impact of different code repair turns, and 2) evaluating the code repair effectiveness on different code error types.

As shown in Figure~\ref{fig:iterations}, the code generation performance is significantly improved during the iteratively repairing. After three turns, the INTERVENOR achieves almost the best performance on HumanEval and HumanEval-X datasets, showing the efficiency of our interactive chain-of-repair mechanism. For a more difficult dataset MBPP, INTERVENOR achieves 30\% improvements and still has some room to achieve further improvements, demonstrating its advantages in dealing with more difficult and realistic coding tasks.

\begin{figure}[t]
    \centering
    \subfigure[AssertionError.] { \label{subfig:error_analysis_1} 
    \includegraphics[width=0.48\linewidth]{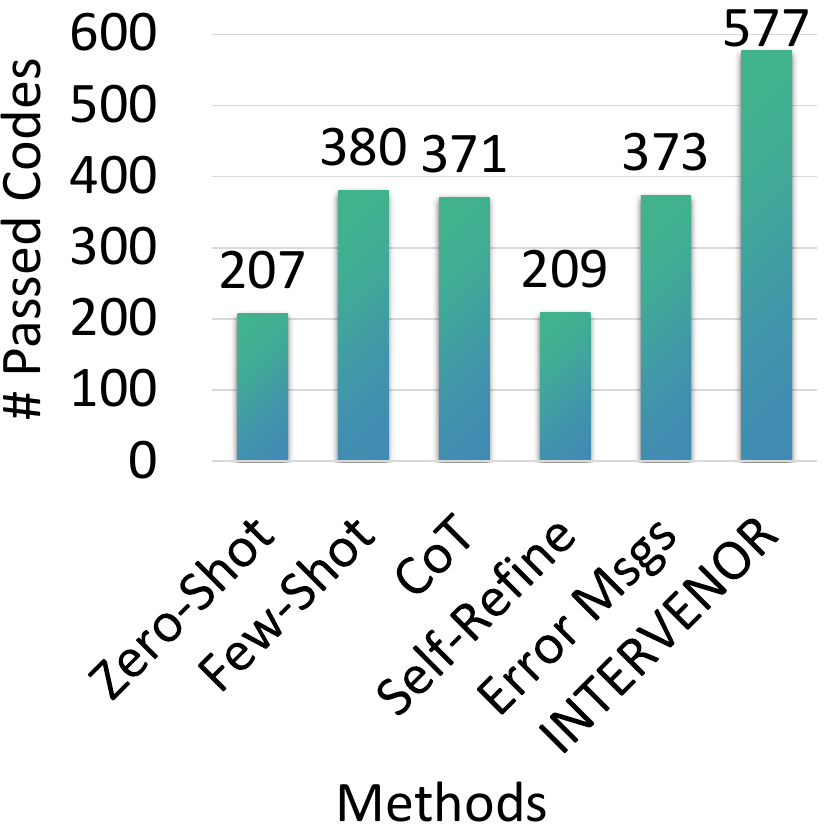}}
    \subfigure[Other Error Types.] { \label{subfig:error_analysis_2} 
    \includegraphics[width=0.48\linewidth]{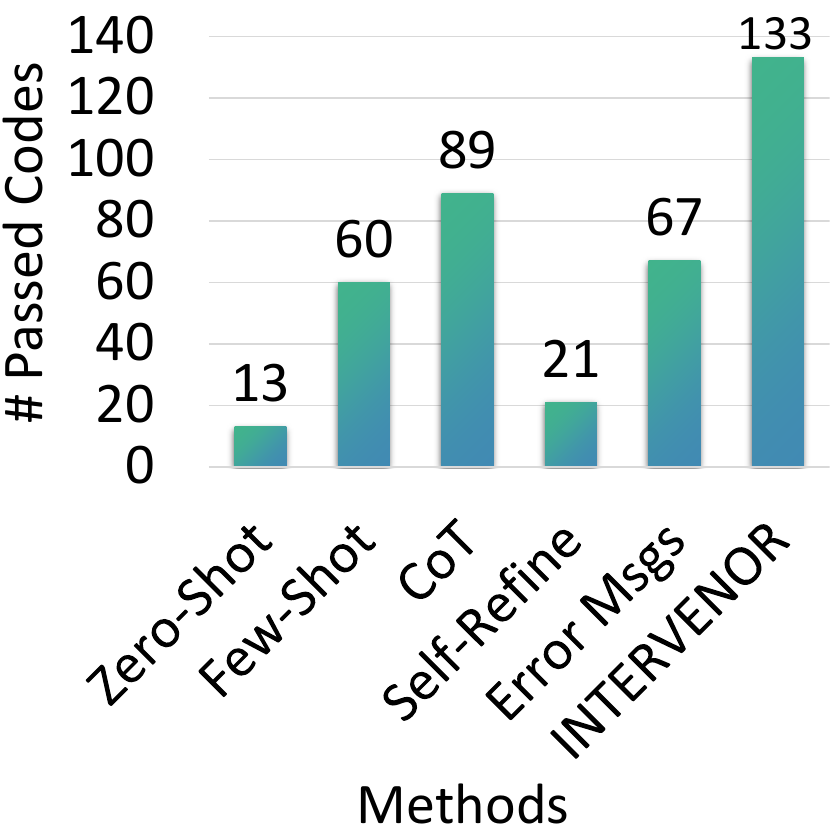}}
    \caption{Code Repair Performance on the CodeError Dataset. We repair the error codes with one single turn. The codes are divided into two groups to evaluate the code repair effectiveness, including Assertion Errors and Others (AttributeError, NameError, RecursionError, SyntacError and TypeError).}
    \label{fig:error_analysis}
\end{figure}

Then we show the code repair effectiveness on different error types in Figure \ref{fig:error_analysis}. We use different prompt methods to stimulate LLMs to repair code errors and show their effectiveness on different types of code errors. Overall, INTERVENOR doubles the number of corrected code examples of baseline models, showing its effectiveness in repairing code errors. On the one hand, INTERVENOR significantly outperforms other methods on the AssertionError repairing task.  It illustrates that INTERVENOR can provide more precise guidance and identify errors with the help of the bug report derived from the failed testing case. On the other hand, INTERVENOR also shows a strong ability to correct other code errors. Our Chain-of-Repair (CoR) method also thrives on the error messages of the compiler, breaks the cognitive inertia of LLMs, and identifies the specific code error for repairing. All these phenomena show that the quality of code execution feedback is critical in repairing codes.

\begin{figure*}[h] \centering
    \includegraphics[scale=0.6]{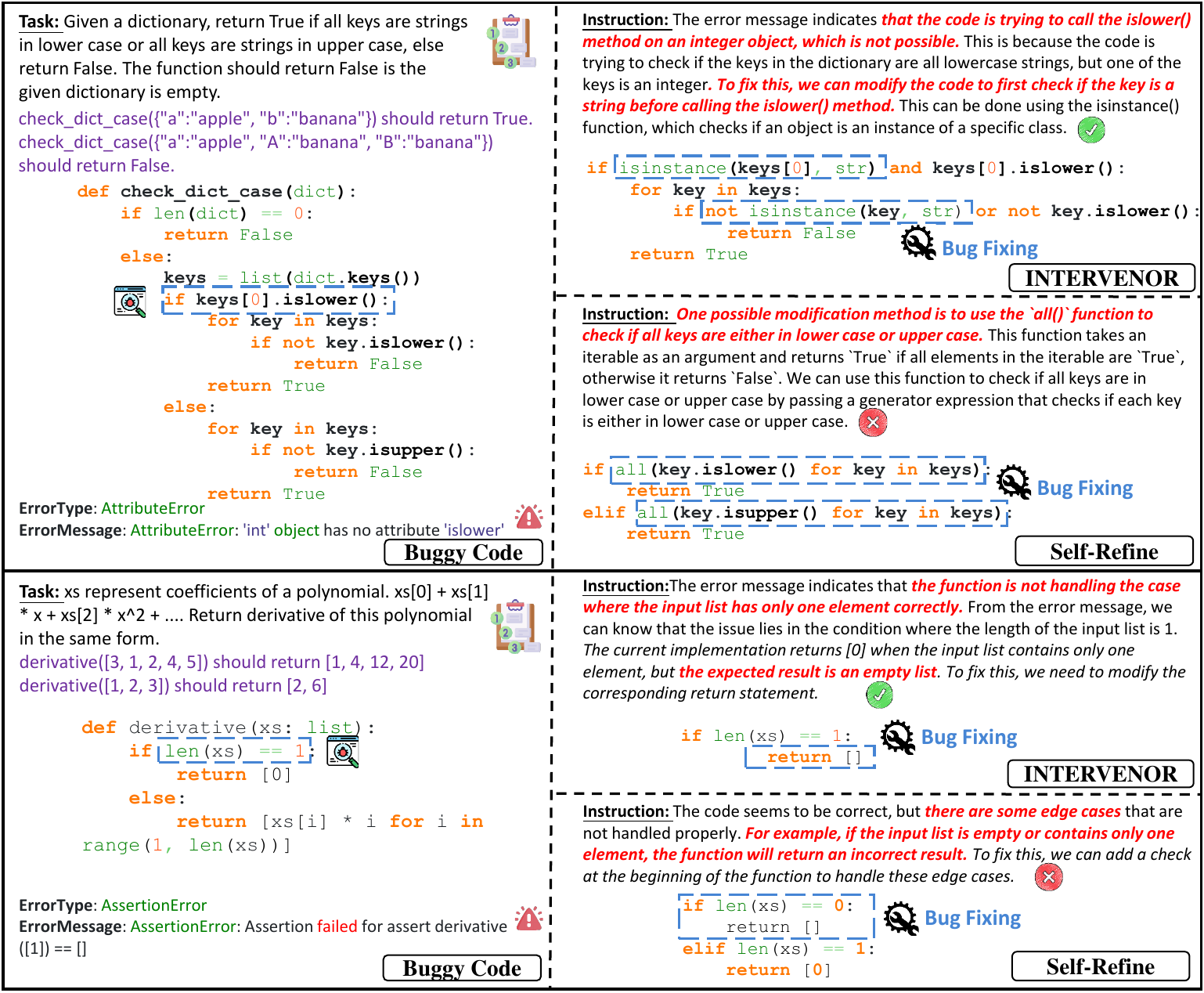}
    \caption{Case Studies. We provide two cases that showcase the effectiveness of the Chain-of-Repair (CoR) generated by INTERVENOR when fixing AttributeError and AssertionError, respectively.} \label{fig:case}
\end{figure*}

\subsection{Case Studies}
Finally, we show two cases in Figure~\ref{fig:case} to demonstrate the effectiveness of INTERVENOR. We compare INTERVENOR with Self-Refine~\cite{madaan2023selfrefine}, which prompts LLMs themselves to execute codes, recognize bugs, and repair codes.

Overall, the feedback from compilers indeed helps to improve the accuracy of repaired code by providing more valuable instructions. In the first case, Self-Refine fails to fix the AttributeError and adds the ``all()'' function in codes. On the contrary, INTERVENOR successfully fixes the code, showing its effectiveness. It accurately analyzes the reason of bugs ``call the islower() method on an integer object, which is not possible'' and also provides a solution by suggesting to ``check if the key is a string before calling the islower() method''. In the second case, the instruction provided by Self-Refine thinks the code seems to be correct and does not offer definite solutions, showing the cognitive inertia in debug--\textit{It is hard to debug the code written by ourselves}. INTERVENOR shows its effectiveness in directly generating the reason for the bug: ``the function is not handling the case where the input list has only one element correctly''. More cases are shown in Appendix~\ref{app:more_cases}.

\section{Conclusion}
This paper proposes INTERVENOR, which regards the feedback of compilers as the INTERVENOR to facilitate the agent collaboration of code error diagnosis and repair. Our experiments show that INTERVENOR outperforms previous code generation/repair models by generating accurate code repair solutions and instructions.

\section*{Limitations}
The effectiveness of INTERVENOR heavily depends on the capabilities of the foundation model, necessitating its proficiency in code error analysis and repair. The efficacy of INTERVENOR may be constrained when using less proficient LLMs to build the agents. Furthermore, in cases where codes exhibit solely logical errors, the compiler relies on test cases to furnish feedback. The quality and coverage of test cases are crucial in this scenario. Inadequate or poorly designed test cases may not be sufficient to detect or help rectify these logical errors effectively. Hence, the efficacy of INTERVENOR is intricately linked to the robustness and comprehensiveness of the test cases.

\section*{Ethics Statement}
In conducting this research and presenting the findings in this paper, we have upheld rigorous ethical standards throughout the entirety of the process. Our experiments are conducted on commonly used datasets from previous works. Additionally, for our newly constructed dataset, we have also reached out to relevant platforms, and they have granted us permission to use the data for scientific research purposes. Besides, the code submitted to the Atcoder website has undergone stringent ethical review, with very few instances where ethical concerns arise. Furthermore, despite the risk of large language models (LLMs) generating toxic data, by providing clear role-playing instructions and only allowing LLMs to complete code-related tasks, we believe the model's output will not produce harmful content.

\section*{Acknowledgments}
Hanbin Wang and Zhenghao Liu are supported by the Natural Science Foundation of China under Grant (No. 62206042 and No. 62272093), the Joint Funds of Natural Science Foundation of Liaoning Province (No. 2023-MSBA-081), and the Fundamental Research Funds for the Central Universities under Grant (No. N2416012). Ge Yu is supported by the Natural Science Foundation of China under Grant (No. 62137001).
%Acknowledgements
% \balance
\bibliography{custom}
\clearpage
\newpage
\appendix

\section{Appendix}

\subsection{License}
For all datasets in our experiments, HumanEval uses the MIT License, MBPP uses the CC-BY-4.0 License, and HumanEval-X uses the Apache License 2.0. All of these licenses allow their data for academic use.

\subsection{Code Translation Results}\label{app:code_trans}
As shown in Table~\ref{tab:overall_code_trans}, we present more detailed evaluation results on the code translation task. 

In general, INTERVENOR shows the best performance across all twelve cross-language code translation tasks, demonstrating its ability to understand and translate codes. Compared with GPT-3.5, INTERVENOR improves its performance on all code translation tasks, especially on the tasks that translate other languages into Python. This indicates that INTERVENOR is more effective in generating an effective code repair chain according to the bug reports from the Python interpreter. Such a phenomenon aligns with human intuition during the debugging process, namely that bugs in Python code are more easily resolved.

% \subsection{XXXX}\label{app:code_trans}

\subsection{More Details of the CodeError Benchmark}\label{app:codeerror}
In our experiments, we build the CodeError benchmark to evaluate the code repair capabilities of LLMs, which facilitates the research on the code repair task. The CodeError dataset is collected from basic programming problems, data analysis problems, and programming competition problems. In this section, we show more detailed data information about the CodeError benchmark. 

\begin{figure*}[t] \centering
    \includegraphics[width=1.0\textwidth]{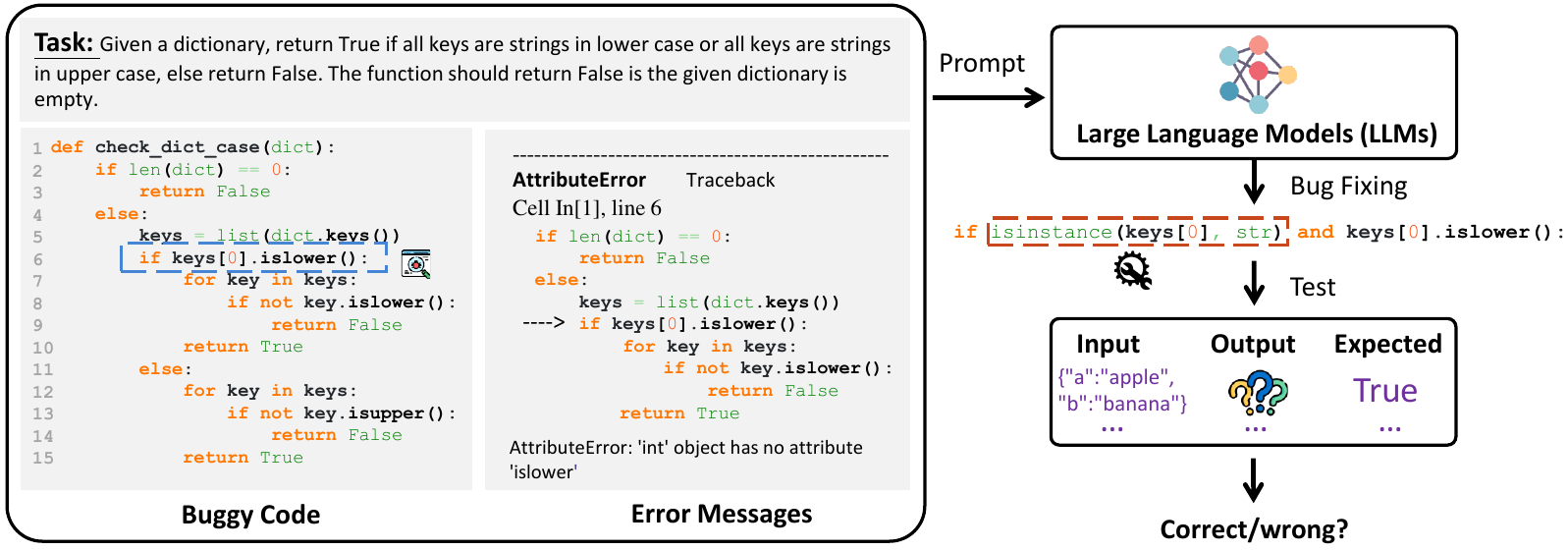}
    \caption{An Example of the CodeError Dataset. The CodeError benchmark asks LLMs to fix buggy codes and then evaluate whether the fixed code meets the requirements specified in the task description and passes all test cases. During the process of code repair, LLMs can utilize error messages to repair codes.} \label{fig:benchmark}
\end{figure*}

\begin{table}[h]
\centering
\small
\resizebox{0.48\textwidth}{!}{
\begin{tabular}{l|l|cccc}
\hline
\multirow{2}{*}{\textbf{Source}}                    & \multirow{2}{*}{\textbf{Model}} & \multicolumn{4}{c}{\textbf{Target}}                          \\
                                     &                                 & \textbf{Python} & \textbf{C++}  & \textbf{Java} & \textbf{JS} \\ \hline
\multirow{4}{*}{\textbf{Python}}     & CodeGen               & -               & 35.9          & 29.3          & 43.4               \\
                                     & CodeGeeX                 & -               & 34.2          & 42.0          & 34.8                \\
                                     & GPT-3.5               & -               & 62.8          & 70.7          & 82.3                \\
                                     & INTERVENOR                            & -               & \textbf{67.7} & \textbf{75.1} & \textbf{87.8}       \\ \hline
\multirow{4}{*}{\textbf{C++}}        & CodeGen               & 33.8            & -             & 43.2          & 54.5                \\
                                     & CodeGeeX                 & 62.8            & -             & 71.7          & 50.8               \\
                                     & GPT-3.5               & 81.1            & -             & 89.6          & 82.3                \\
                                     & INTERVENOR                            & \textbf{90.2}   & -             & \textbf{92.1} & \textbf{86.1}       \\ \hline
\multirow{4}{*}{\textbf{Java}}       & CodeGen               & 52.7           & 41.4          & -             & 57.7                \\
                                     & CodeGeeX               &75.0   & 49.7          & -             & 50.0                \\
                                     & GPT-3.5               & 89.6            & 75.6          & -             & 89.1                \\
                                     & INTERVENOR                            & \textbf{92.1}            & \textbf{79.3} & -             & \textbf{90.9}       \\ \hline
\multirow{4}{*}{\textbf{JS}} & CodeGen               & 35.5            & 35.4          & 33.8           & -                   \\
                                     & CodeGeeX                 & 67.7   & 46.9          & 56.6          & -                   \\
                                     & GPT-3.5               & 82.3            & 76.2          & 84.8          & -                   \\
                                     & INTERVENOR                            & \textbf{87.2}            & \textbf{79.9} & \textbf{89.1} & -                   \\ \hline
\end{tabular}}
\caption{Code Translation Performance on Humaneval-X. We evaluate the code translation effectiveness among different program languages, including Python, C++, Java, and JavaScript (JS). We report the results of INTERVENOR, which only repairs codes with a single turn. All evaluation results are evaluated with Pass@$1$.}
\label{tab:overall_code_trans}
\end{table}
\begin{table*}[h]
\small
\centering
\scalebox{0.94}{
\begin{tabular}{llrccccc}
\hline
\textbf{Dataset}       & \textbf{Language}    & \textbf{Size} & \multicolumn{1}{r}{\textbf{Avg.TC}} & \textbf{Error Type} & \textbf{Error Msg} & \textbf{Category}                                                                                \\ \hline
DeepFix~\cite{yasunaga2021break}       & C           & 6,971            & \ding{55}                        & CE Only    & \ding{55}     & Basic                                                                                     \\
Review4Repair~\cite{huq2022review4repair} & Java        & 2,961            & \ding{55}                         & All          & \ding{55}     & Basic                                                                                      \\
Bug2Fix~\cite{lu2021codexglue}       & Java        & 5,835      & \ding{55}                        & All          & \ding{55}     & Basic                                                                                      \\
Github-Python~\cite{yasunaga2021break} & Python      & 15k             & \ding{55}                        & CE Only    & \ding{55}     & Basic                                                                                       \\
FixEval~\cite{haque2023fixeval}       & Java/Python & 43k/243k       & 25.0                                  & All          & \ding{55}     & Competition                                                                                      \\ \hline
CodeError (Ours)     & Python      & 4,463            & 9.0                                  & All          & \ding{51}           & \begin{tabular}[c]{@{}c@{}}Basic \\ Data Analysis\\ Competition\end{tabular} \\ \hline
\end{tabular}
}
\caption{A Comparison between CodeError and Other Code Repair Benchmarks. Size only represents the size of the test set. Avg.TC indicates the average number of test cases per problem. CE indicates compilation errors (e.g., SyntaxError). Error Msg indicates whether the buggy code contains detailed error information, such as the line of incorrect code, the reasons for the error, etc. Category represents the scope covered by the buggy code, and CodeError covers basic programming problems, data analysis problems, and programming competition problems.}
\label{tab:comparison}
\end{table*}

\begin{table}[h]
\small
\centering
\scalebox{0.948}{
\begin{tabular}{l|rrr|r}
\hline
\textbf{}         & \textbf{Basic} & \textbf{Comp} & \textbf{DA} & \textbf{Total/Avg.} \\ \hline
Problem           & 326            & 3,888          & 249         & 4,463               \\
AssertionError    & 236            & 2,949          & 52          & 3,237                \\
NameError         & 22             & 62            & 32          & 116                 \\
TypeError         & 39             & 91            & 53          & 183                 \\
IndexError        & 1              & 92            & 8           & 101                 \\
ValueError        & 2              & 229           & 44          & 275                 \\
SyntaxError       & 11             & 375           & 4           & 390                 \\
Other Errors      & 15             & 90            & 57          & 162                 \\ \hline
Avg. Problem Words & 10             & 47             & 140            & 49                   \\
Avg. Buggy Code    & 21             & 34             & 2           &   31                 \\
Avg. Test Cases    & 4              & 10             & 1.6         & 9                   \\ \hline
\end{tabular}
}
\caption{Data Statistics of CodeError. We calculate the average word count per problem, the average number of lines in buggy code, and the average number of test cases per problem. Basic, Comp, and DA represent basic programming problems, programming competition problems, and data analysis problems, respectively.}
\label{tab:data_statistics}
\end{table}

\begin{figure*}[t] \centering
    \includegraphics[scale=0.60]{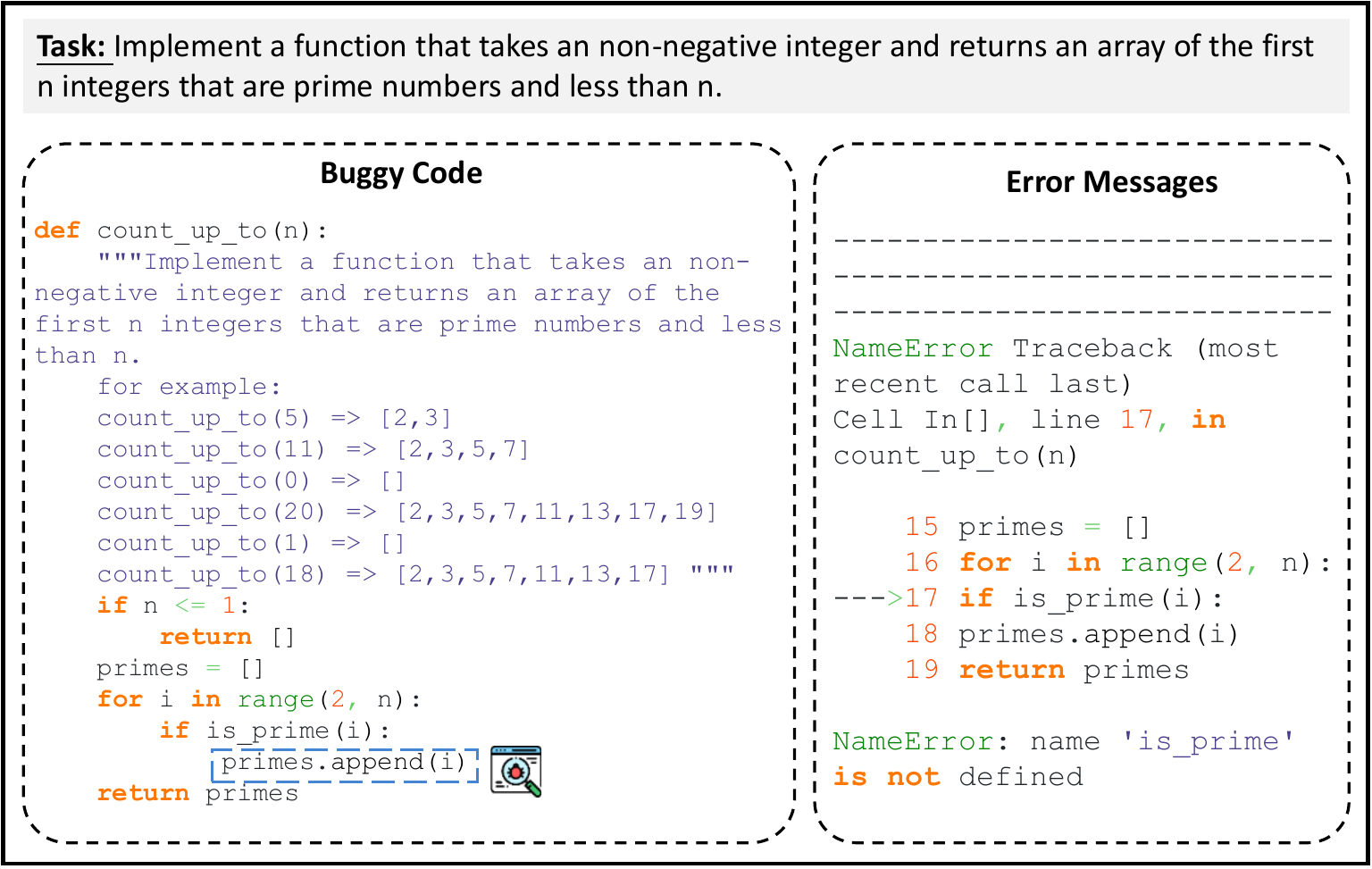}
    \caption{An Example from the Basic Programming Problems.} \label{fig:benchmark_cases_basic}
\end{figure*}

\begin{figure*}[t] \centering
    \includegraphics[scale=0.60]{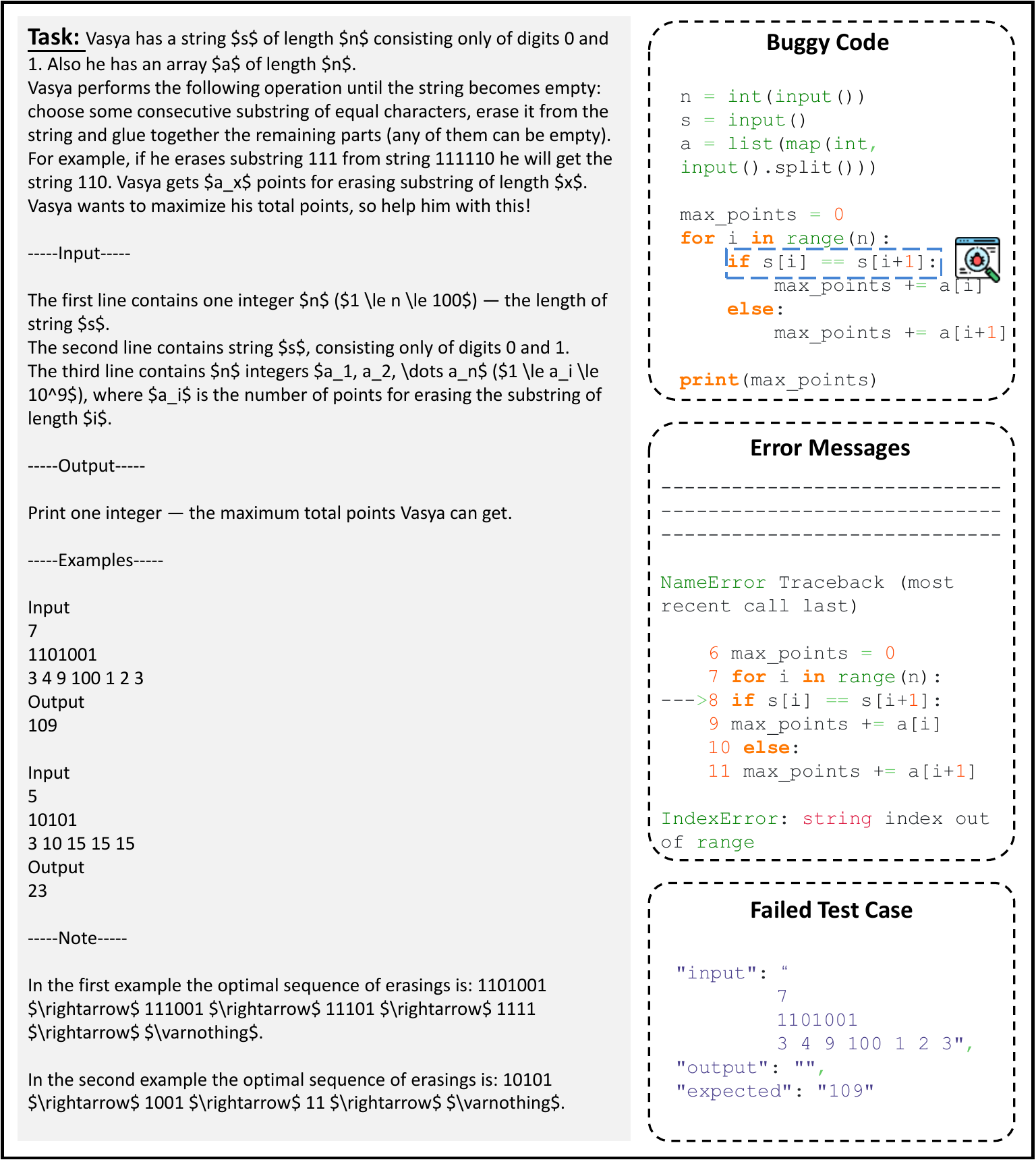}
    \caption{An Example from the Programming Competition Problems. The error information includes the line of code where the error occurs and the test case that failed.} \label{fig:benchmark_cases_comp}
\end{figure*}

\begin{figure*}[t] \centering
    \includegraphics[scale=0.60]{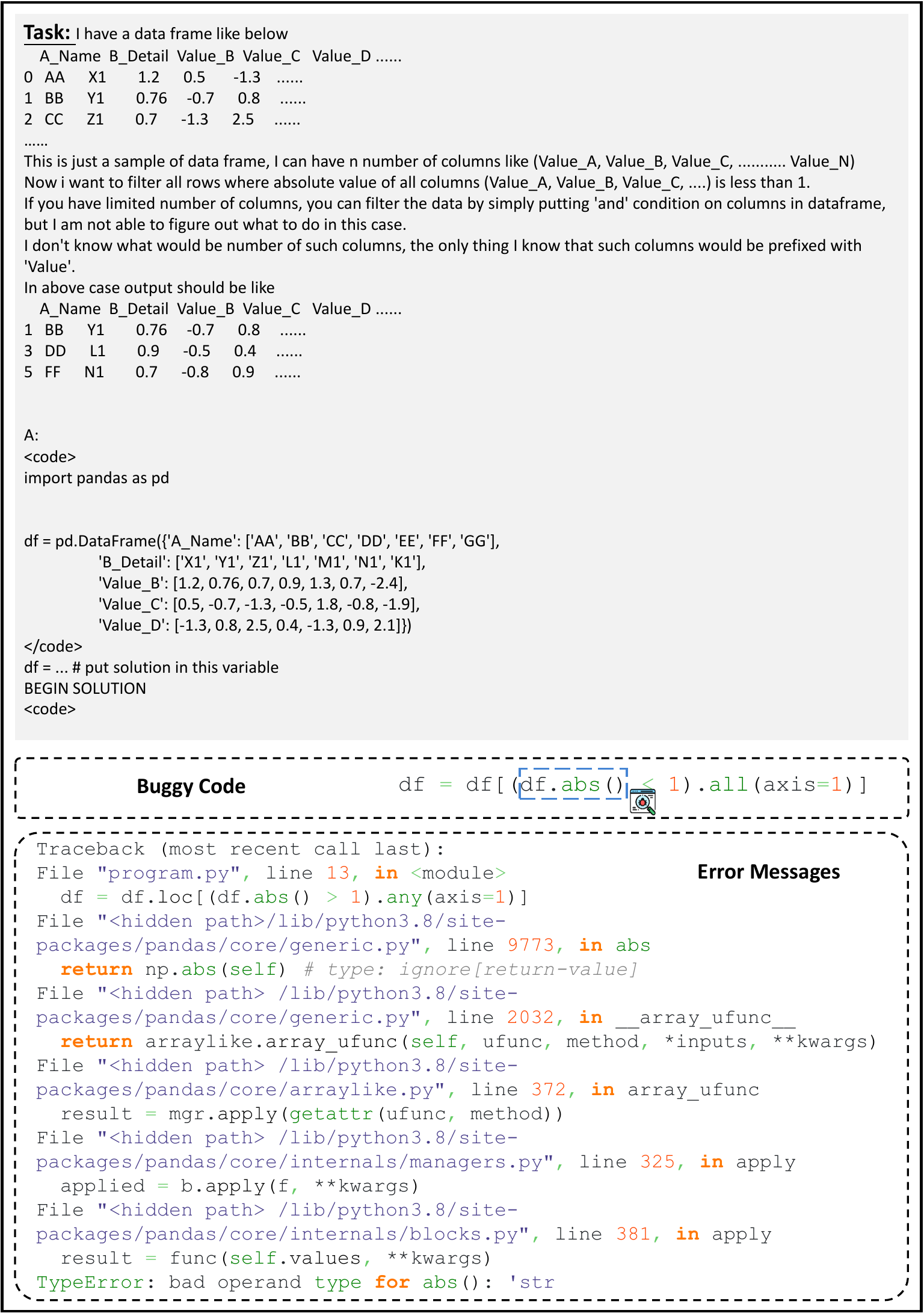}
    \caption{An Example from the Data Analysis Problem. We use the completion mode in DS-1000.} \label{fig:benchmark_cases_da}
\end{figure*}

\begin{table}[h]
\small
\begin{tabular}{p{7.3cm}}
\hline
\textbf{AssertionError} is an exception that is raised when an assert statement fails. In this paper, an AssertionError indicates that the code can run correctly but fails to pass certain test cases, suggesting the presence of potential logic errors that require further investigation and resolution.
\\ \\
\textbf{NameError} is an exception that is raised when a local or global name is not found. This error occurs when you try to access a variable or a function that is not defined or is not in the current scope.
 \\ \\
\textbf{TypeError} is an exception that is raised when an operation or function is applied to an object of an inappropriate type. This typically occurs when you try to perform an operation that is not supported for the type of data you are working with.
  \\ \\
\textbf{IndexError} is an exception that is raised when you try to access an index that does not exist in a list, tuple, or any other sequence. This typically happens when you attempt to access an index that is outside the range of the sequence.
   \\ \\
\textbf{ValueError} is an exception that is raised when a built-in operation or function receives an argument that has the right type but an inappropriate value. Essentially, this error occurs when a function receives an argument of the correct type, but the value of the argument is not appropriate for the operation.
    \\ \\
\textbf{SyntaxError} is an exception that is raised when there is an error in the syntax of your code. This can happen due to various reasons, such as missing parentheses, invalid keywords, or incorrect indentation.
\\ \hline
\end{tabular}
\caption{Common Code Errors and Their Descriptions.}
\label{tab:error_description}
\end{table}
\begin{table*}[h]
\centering
\small
\begin{tabular}{ccccc}
\hline
\textbf{Turns} & \textbf{Self-Debug} & \textbf{Self-Collaboration} & \textbf{Self-Refine} & \textbf{INTERVENOR (Ours)} \\ \hline
1              & 3                   & \textgreater 3                     & 2                    & \textbf{2}                 \\
2              & 6                   & \textgreater 6              & 4                    & \textbf{4}                 \\
3              & 9                   & \textgreater 9              & 6                    & \textbf{6}                 \\ \hline
\end{tabular}
\caption{A Comparison of Computational Overhead Between INTERVENOR and Other Iterative Self-Refinement Baselines.}
\label{tab:computational_overhead}
\end{table*}
\textbf{CodeError Examples.}
As shown in Figure~\ref{fig:benchmark}, each example in CodeError consists of a programming task description, a buggy code snippet, error messages, and test cases to evaluate the code correctness of the repaired codes. If the fixed/generated code meets the requirements specified in the task description and passes all test cases, it proves that the repaired code is correct. During the process of code repair, LLMs need to fix the buggy code and are also able to use the error messages from the buggy code for assistance. We provide more examples that are sampled from the basic programming problems, programming competition problems, and data analysis problems in Figure~\ref{fig:benchmark_cases_basic}, Figure~\ref{fig:benchmark_cases_comp}, and Figure~\ref{fig:benchmark_cases_da}, respectively.

\textbf{Data Collection.}
CodeError is collected from various coding problems, making the dataset diverse and reliable. 

To ensure the diversity of CodeError, we collect basic programming problems, data analysis problems, and programming competition problems from HumanEval, MBPP, DS-1000, APPS, and the programming contest site AtCoder. We conduct code generation experiments using GPT-3.5 on the first four datasets, preserving the generated codes that contain errors. For the programming contest site AtCoder, we crawl real-world user-submitted buggy codes to ensure the reliability of CodeError. The diverse data sources allow us to build a comprehensive and robust dataset to estimate the code repair ability of LLMs.

\textbf{Data Statistics.}
The data statistics of CodeError are shown in Table~\ref{tab:data_statistics}.

The CodeError benchmark contains a total of 4,463 examples, evenly distributed across more than six different error types. These errors range from simple syntax errors to complex logic errors. And we provide detailed descriptions for common error types in Table~\ref{tab:error_description}. The predominant error type is the AssertionError (Logic Error). This phenomenon is quite normal in the real-world coding scenario since the current integrated development environment (IDE) can assist developers in avoiding simple errors such as SyntaxError and NameError but may not help to identify logic errors within the code.
For each example, there are 9 test cases on average to evaluate the correctness of the repaired codes.

\textbf{Dataset Comparison.}
Finally, we compare the CodeError dataset with other code repair benchmarks. The differences are shown in Table~\ref{tab:comparison}.

Similar to FixEval~\cite{haque2023fixeval}, we utilize test cases to assess the functional correctness of the repaired code. Additionally, in comparison to other benchmarks~\cite{yasunaga2021break,huq2022review4repair,lu2021codexglue,haque2023fixeval}, we have two notable features:
\begin{itemize}
    \item We provide detailed error information for erroneous codes, including error code localization, the cause of the error, error type, and more.
    \item CodeError covers a variety of problem types, including basic programming problems, data analysis problems, and programming competition problems, enabling a more comprehensive benchmark to evaluate the code repair capabilities of LLMs.
\end{itemize}

\subsection{Additional Case Studies}\label{app:more_cases}
In this subsection, we sample some cases from CodeError to demonstrate the effectiveness of the Chain-of-Repair (CoR) mechanism.

\begin{figure*}[h] \centering
    \includegraphics[scale=0.61]{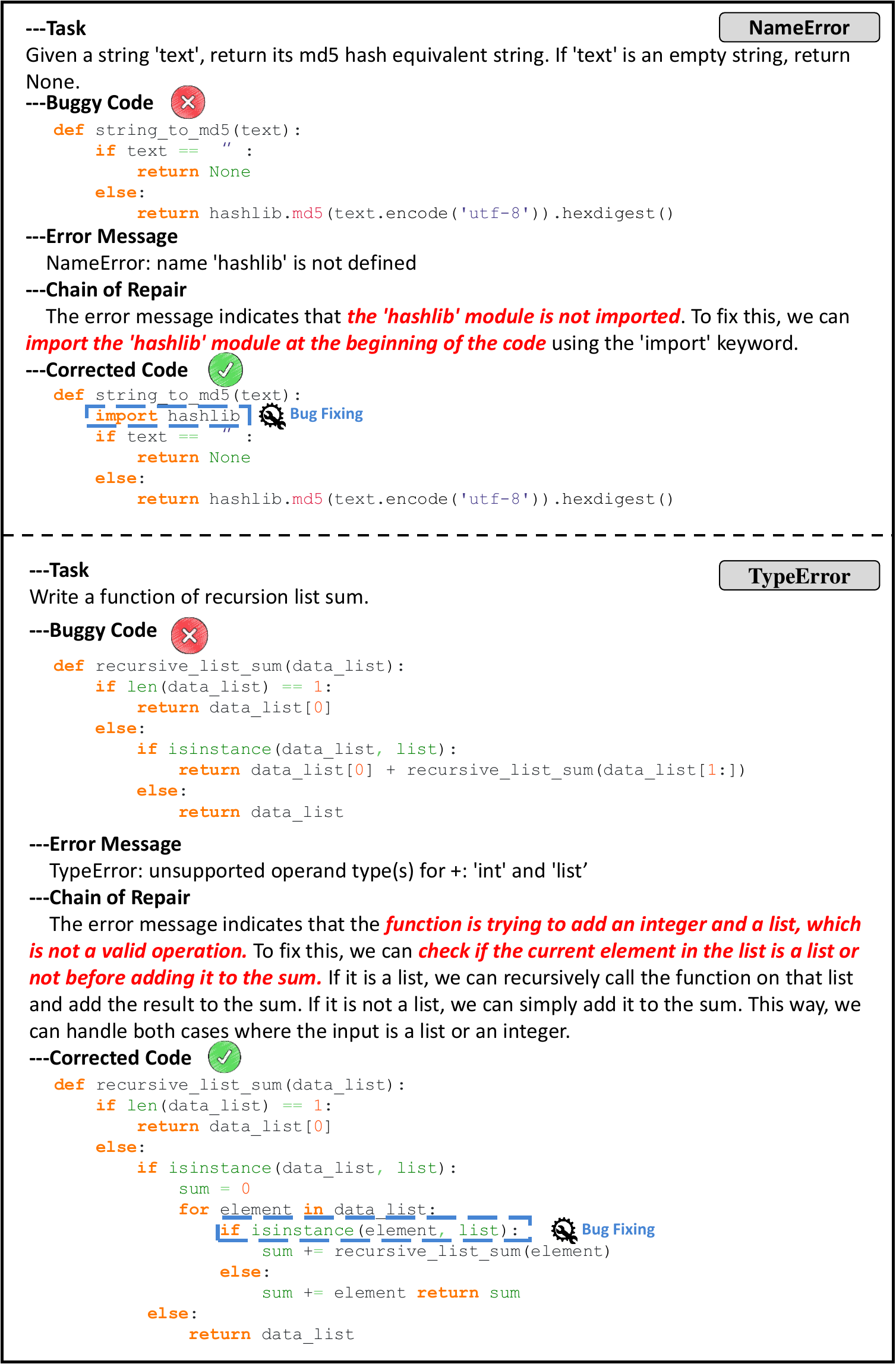}
    \caption{Case Studies. We present two cases from CodeError. The Chain-of-Repair clearly identifies the NameError and TypeError in the buggy code and provides the correct modification methods, which are highlighted in \textcolor{red}{\textbf{red}}.} \label{fig:more_cases3}
\end{figure*}

As shown in Figure~\ref{fig:more_cases3}, in the first case, there is a simple NameError in the buggy code, which indicates that `hashlib' is not defined. We can see that the INTERVENOR recognizes hashlib as a Python package and provides a solution: ``import the `hashlib' module at the beginning of the code''. In the second case, the error in the code is more subtle, involving an operation that should check the data structures of `int' and `list'. The CoR explicitly states that this is a ``valid operation'' and provides a solution: ``check if the current element in the list is a list or not before adding it to the sum''. With the help of accurate instruction, INTERVENOR successfully repairs the code.

\begin{figure*}[h] \centering
    \includegraphics[scale=0.6]{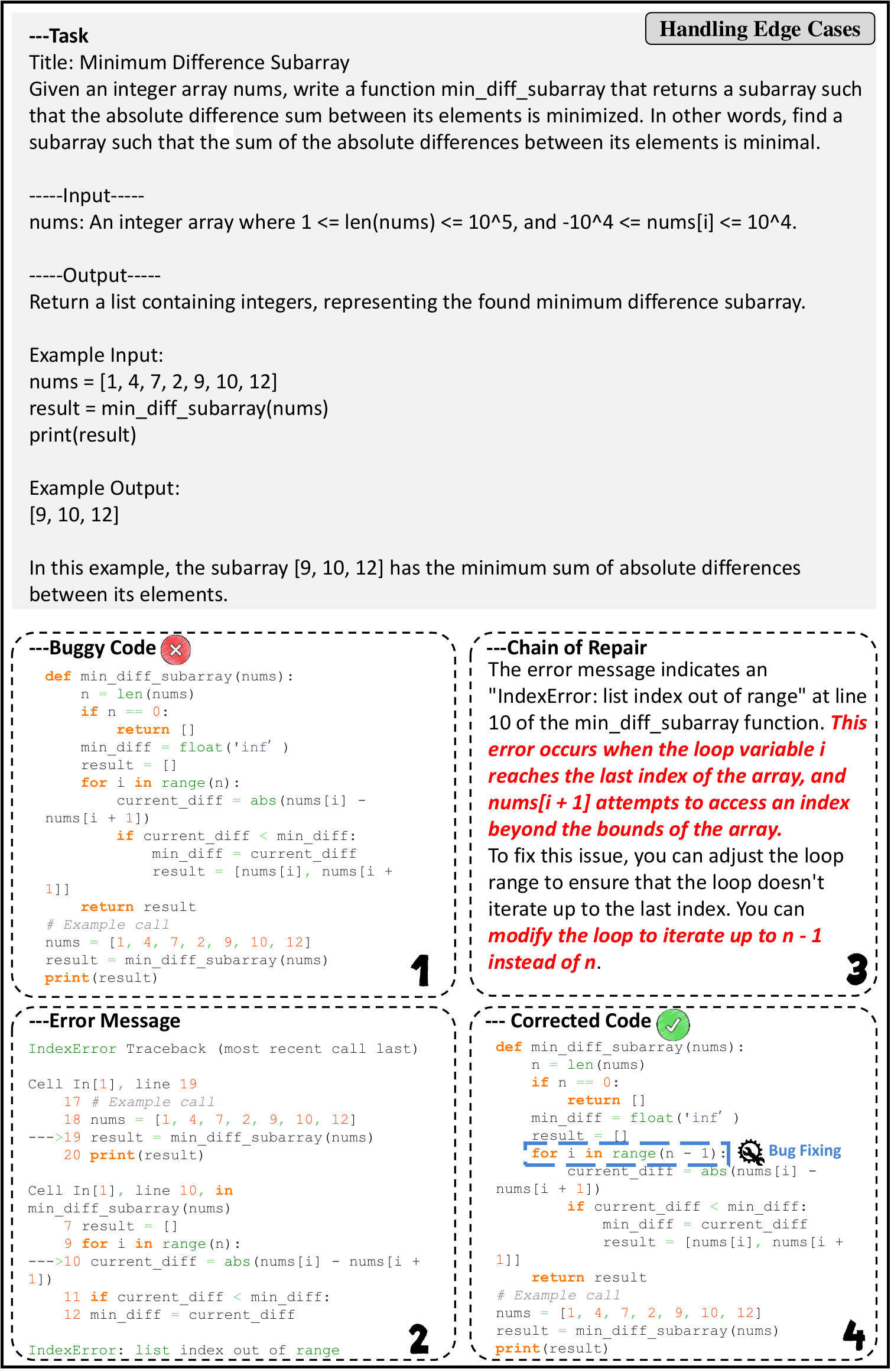}
    \caption{Handling Boundary Issues.} \label{fig:edge_case}
\end{figure*}

\begin{figure*}[h] \centering
    \includegraphics[scale=0.6]{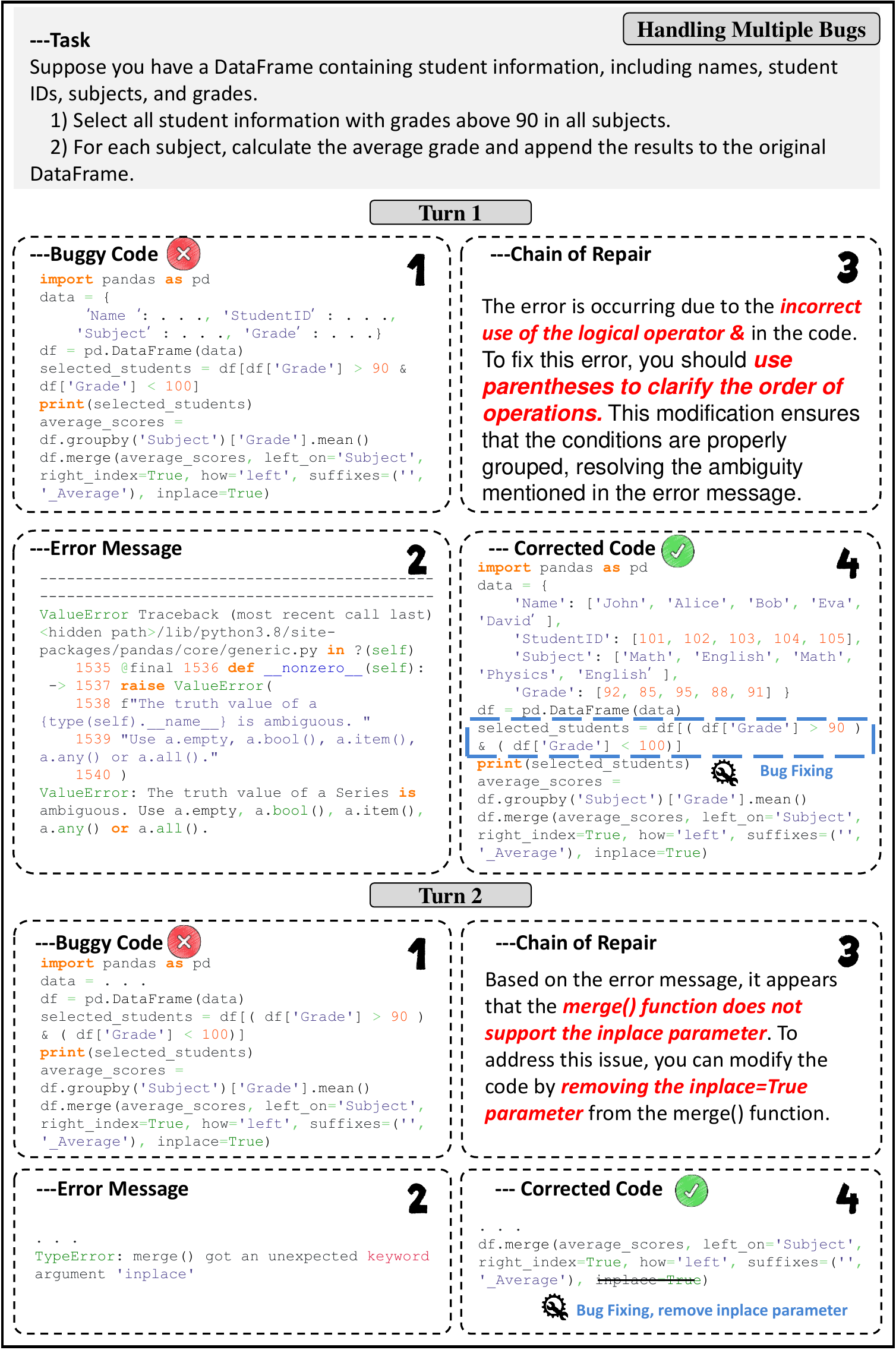}
    \caption{Handling Multiple Bugs.} \label{fig:multi_bugs}
\end{figure*}

Additionally, in Figure~\ref{fig:edge_case}, we demonstrate that INTERVENOR can effectively address boundary issues, which are often quite tricky. INTERVENOR accurately identifies the code error in the loop condition and suggests a modification method: ``modify the loop to iterate up to $n - 1$ instead of $n$''. In Figure~\ref{fig:multi_bugs}, we demonstrate that INTERVENOR can fix multiple error types in the code by iteratively repairing. Over two iterations, INTERVENOR identifies and accurately modifies the ValueError and TypeError in the code. 

All these cases demonstrate the effectiveness of the Chain-of-Repair (CoR) mechanism and highlight the effectiveness of CoR in identifying errors and providing appropriate solutions for code repair.


\begin{figure*}[h] \centering
    \includegraphics[scale=0.61]{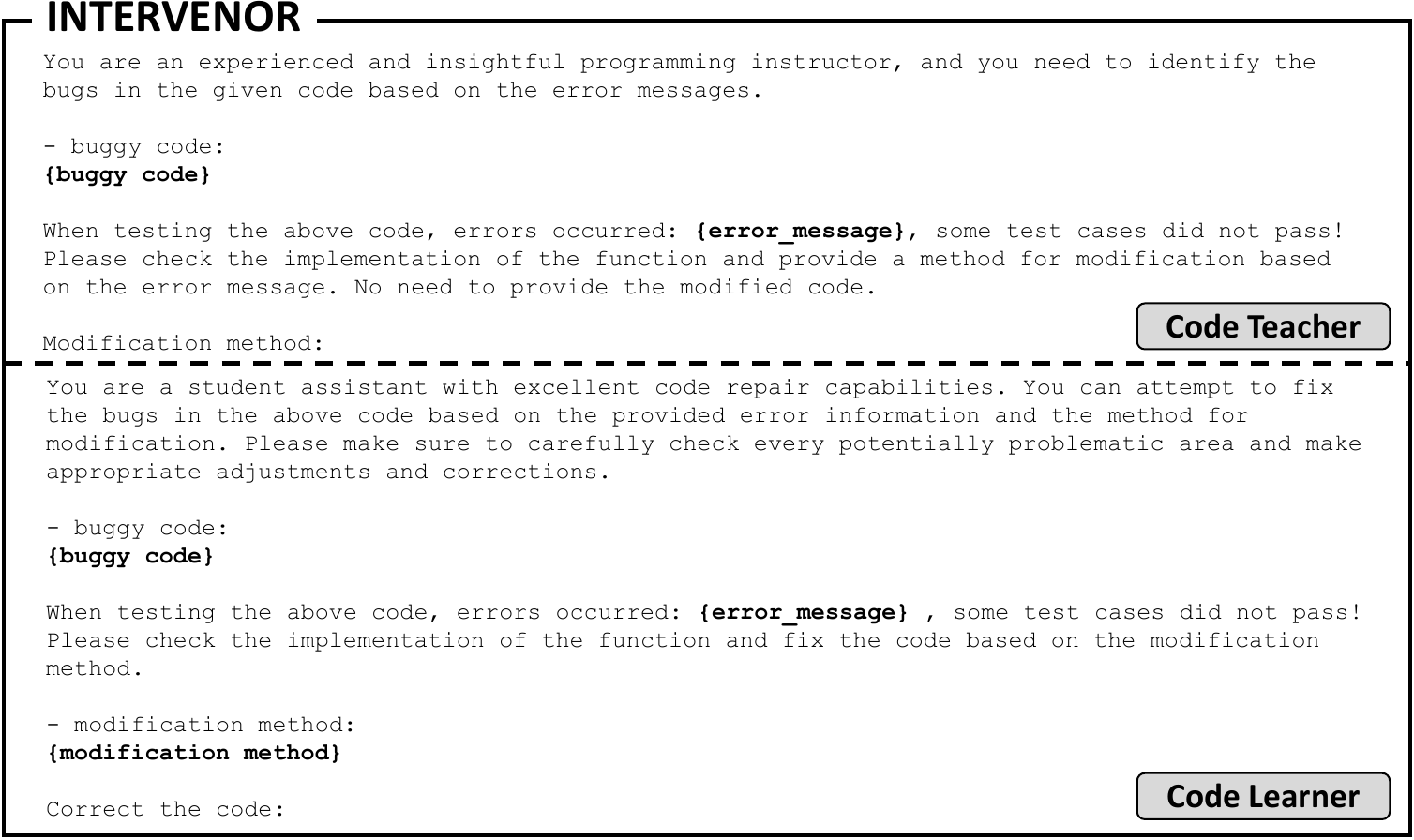}
    \caption{Role Instructions in INTERVENOR. Within INTERVENOR, there are two LLM-based agents \texttt{Code Teacher} and \texttt{Code Learner}. We utilize specific instructions to ensure that they play the correct roles and carry out the intended tasks.} \label{fig:prompt}
\end{figure*}

\begin{figure*}[t] \centering
    \includegraphics[scale=0.61]{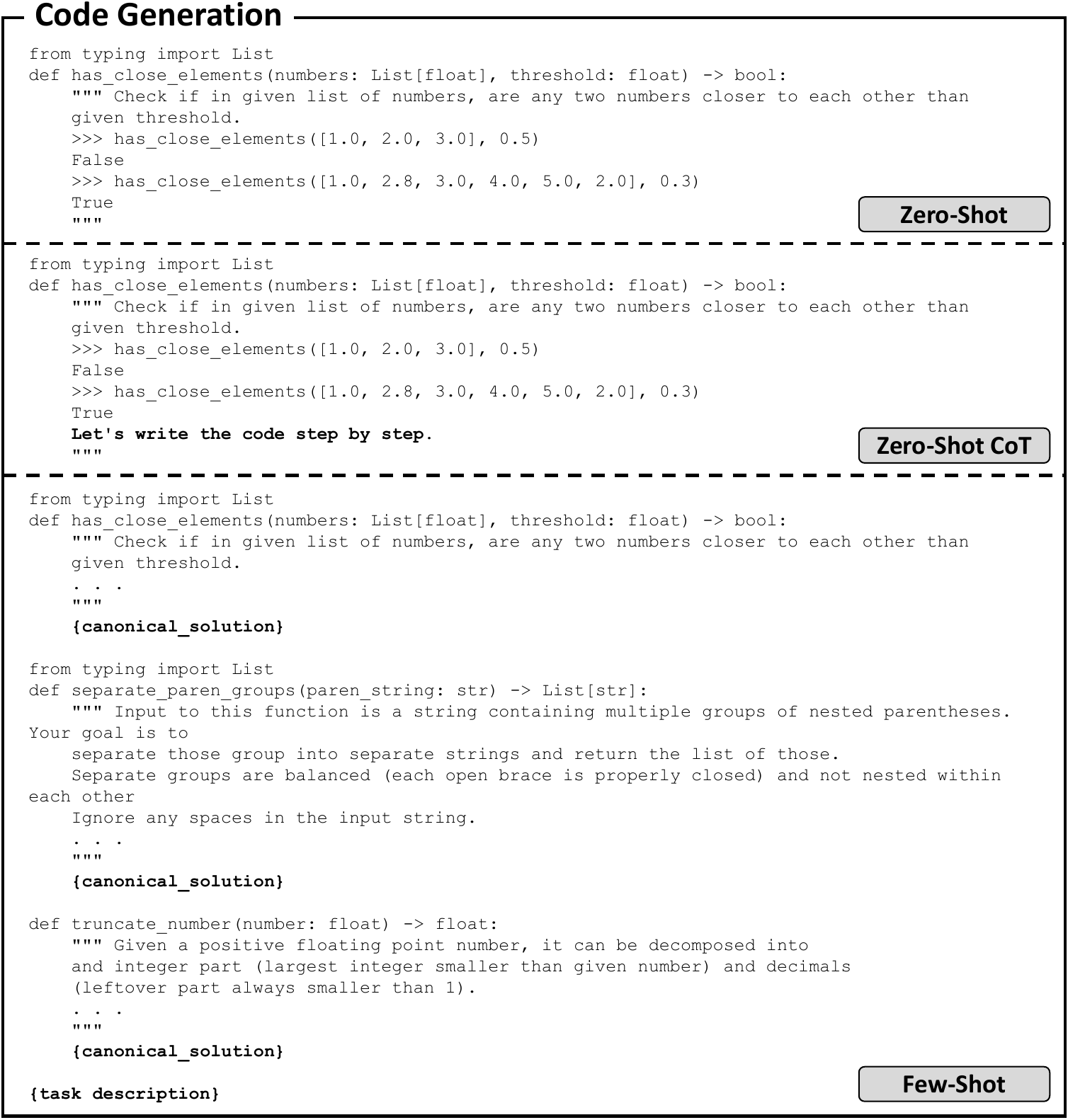}
    \caption{Zero-Shot, Zero-Shot CoT, and Few-Shot Prompts for Code Generation Tasks.} \label{fig:prompt_2}
\end{figure*}

\begin{figure*}[t] \centering
    \includegraphics[scale=0.61]{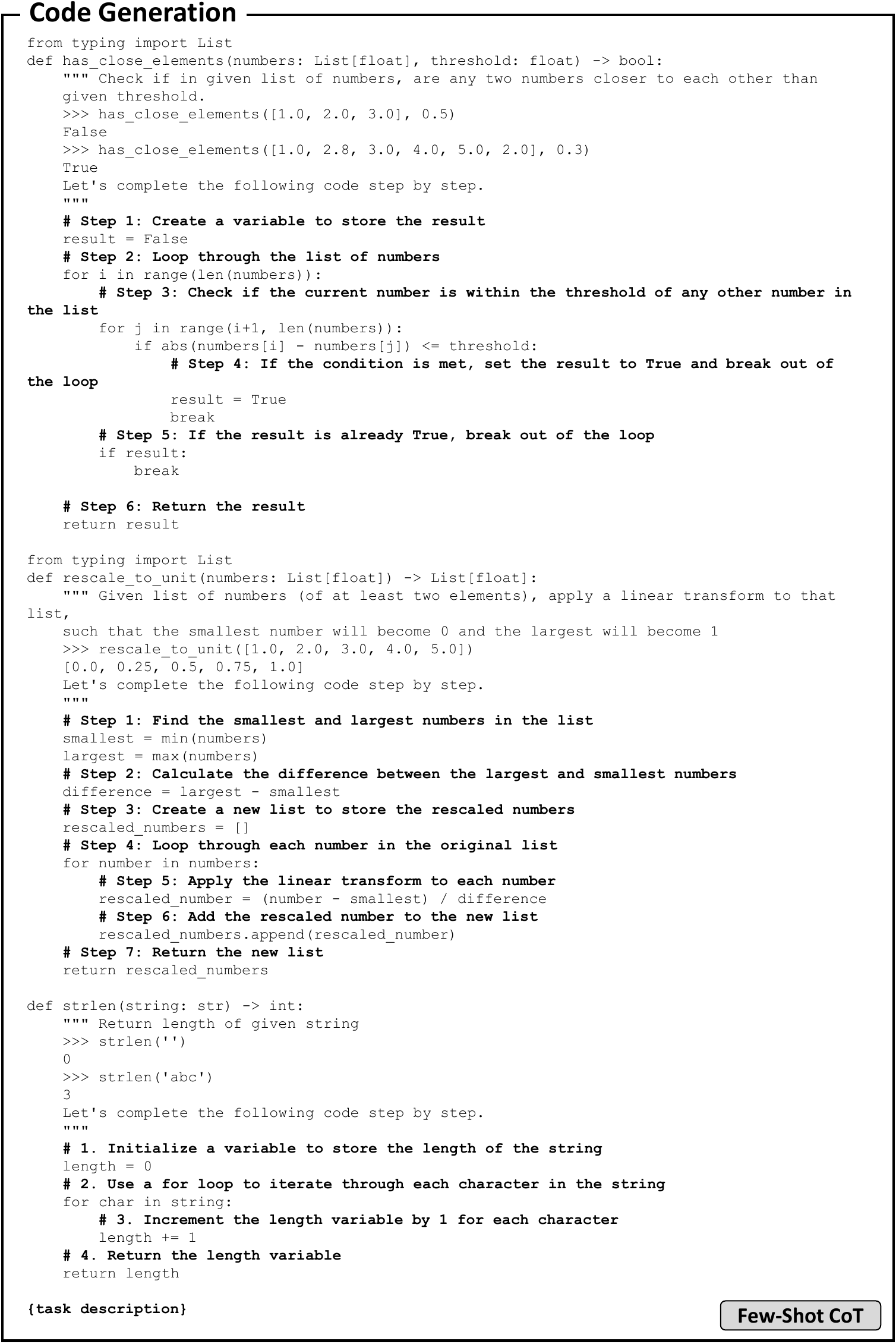}
    \caption{Few-Shot CoT Prompts for Code Generation Tasks.} \label{fig:prompt_3}
\end{figure*}

\begin{figure*}[t] \centering
    \includegraphics[scale=0.61]{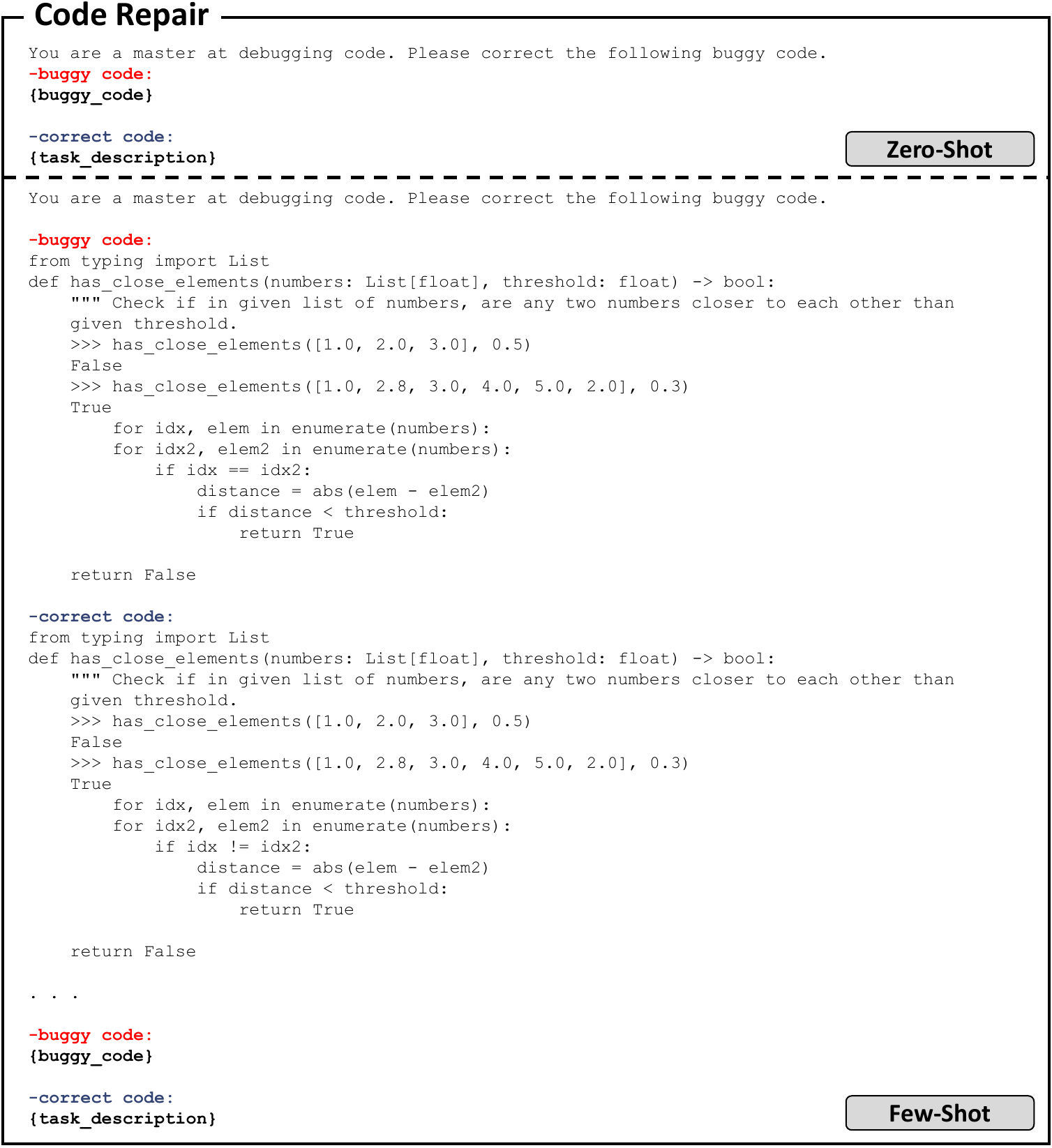}
    \caption{Zero-Shot and Few-Shot Prompts for Code Repair.} \label{fig:prompt_4}
\end{figure*}

\begin{figure*}[t] \centering
    \includegraphics[scale=0.61]{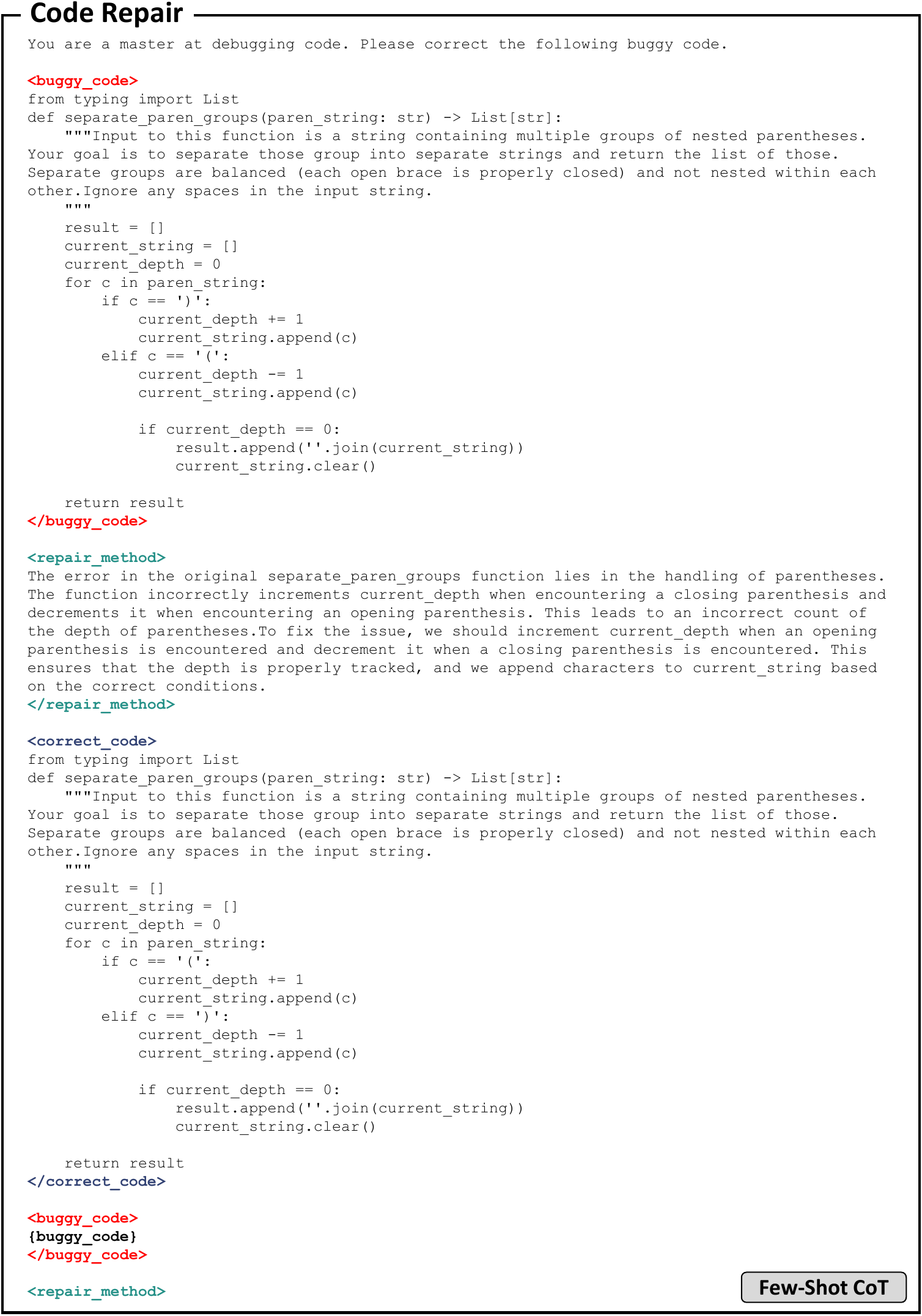}
    \caption{Few-Shot CoT Prompts for Code Repair.} \label{fig:prompt_5}
\end{figure*}

\begin{figure*}[t] \centering
    \includegraphics[scale=0.61]{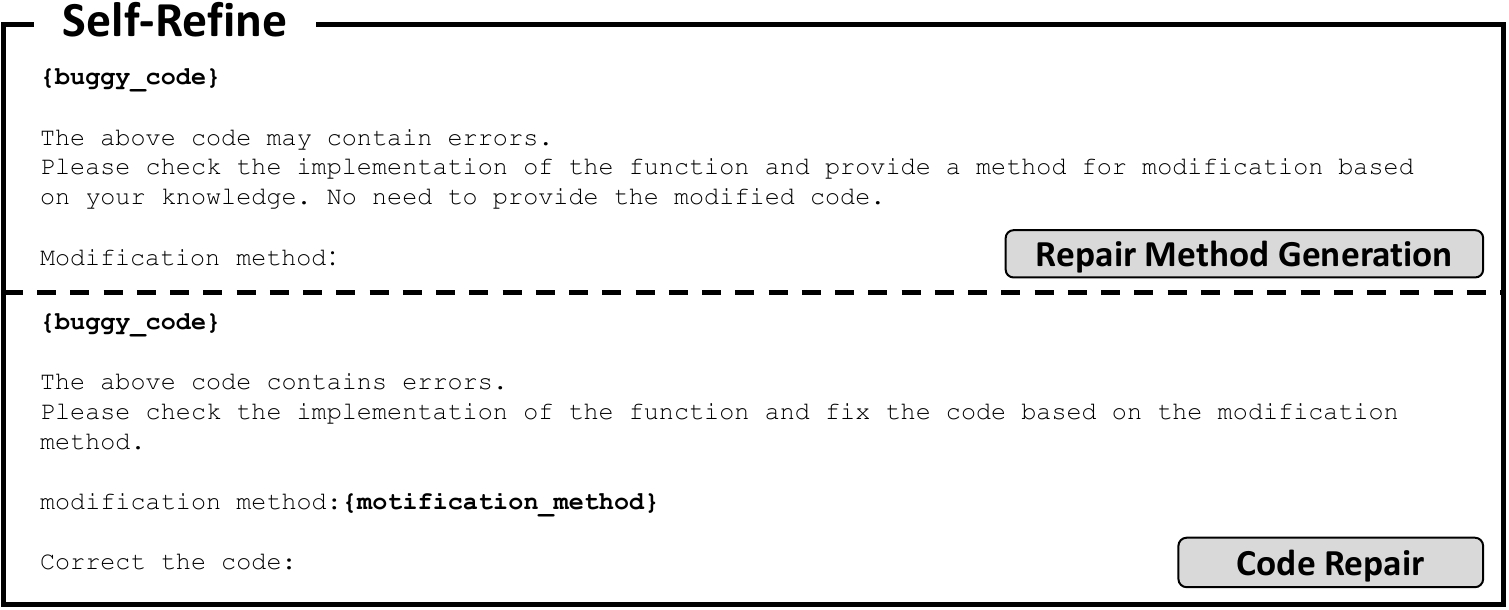}
    \caption{Prompts Used in Self-Refine Model. The instructions for generating code repair plannings and code repair are shown.} \label{fig:prompt_6}
\end{figure*}

\begin{figure*}[t] \centering
    \includegraphics[scale=0.61]{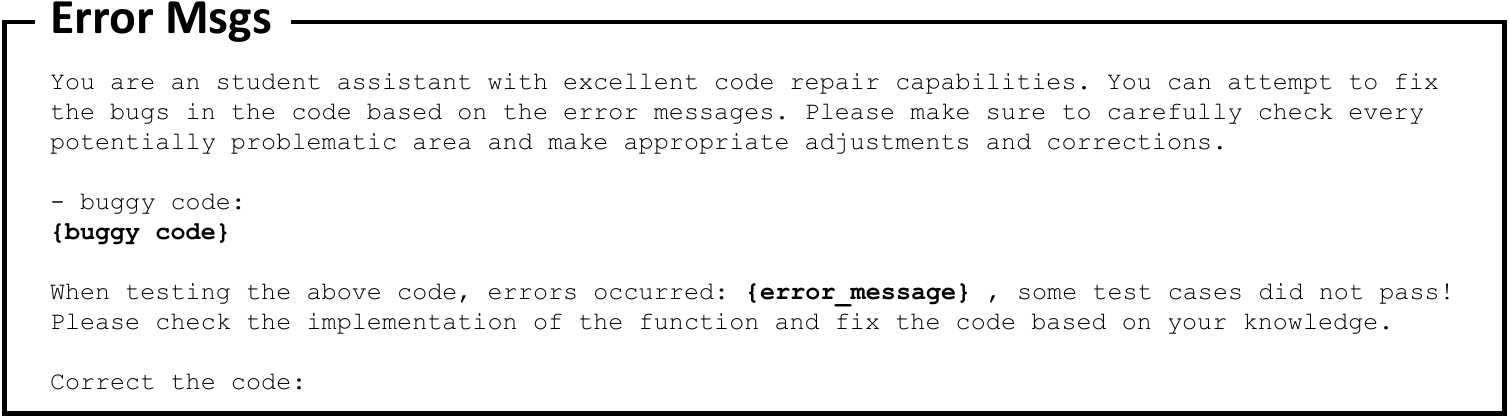}
    \caption{Prompts Used in Error Msgs Model. It conduct code repair by directly using the code error messages.} \label{fig:prompt_7}
\end{figure*}

\subsection{Prompts for Agent Building}\label{app:prompt}
In this subsection, we show the prompts used in our experiments. 

We first show the role-playing instructions used by \texttt{Code Learner} and \texttt{Code Teacher} in Figure~\ref{fig:prompt}. We utilize specific instructions to ensure that the \texttt{Code Teacher} and \texttt{Code Learner} within INTERVENOR play the correct roles and carry out code inspection and code repair. 

Then, we present the prompts used in code generation and code repair. From Figures~\ref{fig:prompt_2} to~\ref{fig:prompt_3}, we show the Zero-Shot, Zero-Shot CoT, Few-Shot, and Few-Shot CoT prompts for code generation tasks. From Figures~\ref{fig:prompt_4} to~\ref{fig:prompt_7}, we show the prompts used in some code repair baseline models, including Zero-Shot, Few-Shot, CoT, Self-Refine, and Error Msgs. Self-Refine asks the LLMs to rethink the errors by themselves, while Error Msgs incorporate the code error messages from compilers and directly use them to guide the code repair process.

\subsection{Computational Overhead}\label{app:computational}
In this subsection, we compare the computational overhead with other iterative self-refinement baselines.

We follow Self-Debug, Self-Collaboration, and Self-Refine to iteratively fix buggy codes. As shown in Table \ref{tab:computational_overhead}, if we iterate for the same number of turns, INTERVENOR calls the API fewer times and brings performance improvements.

\end{document}